\documentclass[12pt]{iopart}
\usepackage[margin=1in]{geometry}
\usepackage{iopams}

\expandafter\let\csname equation*\endcsname\relax
\expandafter\let\csname endequation*\endcsname\relax

\usepackage{amsmath,amssymb}
\usepackage{graphicx}
\usepackage{subfigure}

\begin{document}

\title[Complexified phase spaces, IVRs, and the accuracy of SC propagation]{Complexified phase spaces, initial value representations, and the accuracy of semiclassical propagation}

\author{Gabriel M.~Lando}

\address{Max Planck Institute for the Physics of Complex Systems, N{\"o}thnitzer Stra{\ss}e 38, D-01187 Dresden, Germany}
\ead{lando@pks.mpg.de}
\vspace{10pt}
\begin{indented}
	\item[]June 2020
\end{indented}

\begin{abstract}
	Using phase-space complexification, an Initial Value Representation (IVR) for the semiclassical propagator in position space is obtained as a composition of inverse Segal-Bargmann (S-B) transforms of the semiclassical coherent state propagator. The result is shown to be free of caustic singularities and identical to the Herman-Kluk (H-K) propagator, found ubiquitously in physical and chemical applications. We contrast the theoretical aspects of this particular IVR with the van~Vleck-Gutzwiller (vV-G) propagator and one of its IVRs, often employed in order to evade the non-linear ``root-search'' for trajectories required by vV-G. We demonstrate that bypassing the root-search comes at the price of serious numerical instability for all IVRs except the H-K propagator. We back up our theoretical arguments with comprehensive numerical calculations performed using the homogeneous Kerr system, about which we also unveil some unexpected new phenomena, namely: (1) the observation of a clear mark of half the Ehrenfest's time in semiclassical dynamics; and (2) the accumulation of trajectories around caustics as a function of increasing time (dubbed ``caustic stickiness''). We expect these phenomena to be more general than for the Kerr system alone.      
\end{abstract}


\section*{Introduction}

The fundamental parameter of quantum mechanics, namely Planck's constant or its reduced form $\hbar$, is an extremely small quantity. It did not take long after the Schr\"odinger equation was discovered in order for approximations exploiting this smallness to be developed, especially by van Vleck in his seminal 1928 paper \cite{VanVleck1928}. An arguably surprising consequence of such asymptotic approximations was that the leading order terms in $\hbar$ involved quantities which \emph{made sense} in classical mechanics, such as actions, trajectories and catastrophes \cite{Maslov1981}. This lead to the new field being called \emph{semiclassical mechanics}, evincing that for the first time a link between classical and quantum realms had been devised. Semiclassical objects such as the van~Vleck-Gutzwiller (vV-G) propagator, for instance, established a connection between classical trajectories and quantum superposition, being able to reproduce interference patterns using solely classical data as input. Besides providing quantum mechanics with some geometrical meaning, semiclassical methods are nowadays used to model a plethora of physical and chemical phenomena.

The Herman-Kluk (H-K) propagator is an alternative to vV-G's, \emph{i.e.}~another expression for the semiclassical propagator in the position representation, discovered in the mid 1980s \cite{Herman1984}. Used either raw or as a starting point to other approximation methods, it is the most popular Initial Value Representation (IVR) in chemistry, although the development of benchmark semiclassical methods is truly non-stop (see \cite{Heller1991-2,Rost2004,Campolieti1997,Liu2015,Curchod2018,Micciarelli2019,Gottwald2019,Grossmann2020} for some examples). In addition to providing high-accuracy results for the complicated processes inherent to the chemical sciences, the H-K propagator was also scrutinized by the physical community, who attested for its remarkable accuracy and implementation ease (\emph{e.g.}~\cite{Schoendorff1998,Rost1999,Maitra2000,Zagoya2012,Lando2019-3}). The reason behind such ease is mostly due to the fact that, as all IVRs, it sums over all initial positions and momenta instead of root-searching for specific trajectories, a procedure required by cruder propagators such as vV-G's. However, the H-K propagator has been shown to be more accurate than other IVRs \cite{Kay1993}, but at present a comprehensive analysis of the reasons behind such accuracy appears to be lacking -- especially one that unveils the role played by \emph{caustics}, which are divergences present in most semiclassical propagators, in the approximations.  

In the first part of this manuscript we collect and reinterpret several scattered results in the physical, chemical and mathematical literatures in order to understand the theory behind the H-K propagator. The main step is to identify this propagator as a sequence of two inverse Segal-Bargmann (S-B \footnote{The Segal-Bargmann space has many names, including: Fock, Fock-Bargmann, Fock-Cook, etc. We choose Segal-Bargmann because this is usually the name given to the transformation mapping the position representation to the coherent state one.}) transforms of the semiclassical propagator in the S-B representation (\emph{i.e}~the Weyl-ordered coherent state propagator). This is similar to what was done in \cite{Grossmann1998}, although we do not perform any integrals or use steepest descent methods, relying instead on the usual ``IVR trick'' devised by Miller \cite{Miller2001}. The H-K propagator is then shown to be a mere extension of the semiclassical quantization of linear hamiltonian flows to non-linear ones, the only difference between it and vV-G being the representation chosen. This seemingly small diffrence, however, is responsible for a major contrast between these propagators, as we can then invoke an earlier result that states that the S-B representation does not suffer from caustic singularities as the position one does \cite{Littlejohn1986}. Besides, although we contrast the S-B and position representations, our conclusions generalize to all semiclassical descriptions plagued by caustic singularities, such as the momentum \cite{Littlejohn1992}, mixed \cite{Maslov1981,Kirkwood1933} and even Weyl-Wigner ones \cite{Ozorio1998}.

In this manuscript's second part we proceed to really visualize the impact of caustics in semiclassical propagation. As our toy model we choose the homogeneous Kerr system, which has a 4th-order hamiltonian and has been already investigated in several instances (\emph{e.g.}~\cite{Yurke1986,Averbukh1989,Raul2012,Lando2019}). This system is particularly tractable due to both its classical and quantum dynamics being analytical, but such analyticity does not prevent it from developing a very intricate caustic web, rendering it the perfect laboratory for our purposes. Comparisons are given in all levels of ``purity'', that is, from the semiclassical propagators themselves to quantities obtained by employing them as integral kernels, and it is found that their contrast decreases with each performed integral. In particular, an accumulation of defects in the semiclassical propagator is seen due to caustics in the position representation, while for the S-B representation it is sometimes hard to even separate between the exact quantum result and its semiclassical approximation. For wave functions and autocorrelations, it is seen that caustics are responsible for normalization loss and oscillatory errors. To show this is not due to the vV-G propagator being a sum and H-K an IVR, we also provide comparisons between results obtained from an IVR based on position representation. We then see, as predicted in the first part of the manuscript, that the a collateral effect of avoiding the root-search in the vV-G propagator results in its IVR requiring much denser integration grids in order to achieve equivalent results. We also identify a possibly new phenomenon, in which the crossings of trajectories and caustics start to take longer as time increases, dubbing it ``caustic stickiness''. When generalizable, it might be a fundamental mechanism behind the long-time failure of semiclassical propagators based on representations that contain caustics. \\

The first part of the manuscript is composed of its initial three sections, and the second part deals exclusively with Sec.~\ref{sec:kerr}. The organization of the sections is now in order. In Sec.~\ref{sec:comp} we briefly review the theory of phase space complexification, differentiating it from truly complex phase spaces. Sec.~\ref{sec:lin} deals mainly with 1-parameter groups of symplectic matrices and their quantization in position and S-B representations, and it is here the problem of caustics and the solution provided by the complexified variables are discussed. In Sec~\ref{sec:semi} we review IVRs and generalize the linear theory of Sec.~\ref{sec:lin} to the semiclassical realm, discussing the consequences of using IVR techniques in representations with and without caustics. We then move on to the numerical analysis performed in Sec.~\ref{sec:kerr}, in which the points raised in the first part are visualized. A brief discussion of our results is then included in Sec.~\ref{sec:disc}, and we finish the manuscript with the conclusions of Sec.~\ref{sec:conc}. Three appendices are included: in \ref{App:A} we reproduce a proof of the non-singularity of an important matrix, which happens to be directly responsible for the S-B representation having no caustics; \ref{App:B} exposes a bit of the mathematics of ``Miller's trick'', \emph{i.e.}~the substitution used to obtain IVRs from raw propagators; and finally \ref{App:C} shows how the root-search for the Kerr system was numerically implemented in Sec.~\ref{sec:kerr}.

\section{Complexification}\label{sec:comp}

In this section we present some properties of a particular complexification of $\mathbb{R}^{2n}$, together with its effect on relevant classical objects such as generating functions and canonical forms. The following conventions are used throughout the manuscript:

\begin{itemize}
	\item Vectors are bold, like $\mathbf{p}$ and $\bzeta$, and matrices are upper case, like $\mathcal{M}$, $A$ and $\Gamma$. Latin and Greek scripts stand for real and complex quantities, respectively.
	\item The space $\mathbb{R}^{2n}$ is decomposed as a direct product of positions and momenta, for which we use condensed coordinates organized as 
	\begin{eqnarray}
		(q_1, \dots, q_n,p_1,\dots,p_n) &= (\mathbf{q},\mathbf{p}) \in \mathbb{R}^n \oplus \mathbb{R}^n \sim \mathbb{R}^{2n} \\
		(\zeta_1, \dots, \zeta_n,\zeta^*_1,\dots,\zeta^*_n) &= (\bzeta,\bzeta^*) \in \mathbb{C}^n \oplus \mathbb{C}^n \sim \mathbb{C}^{2n} \, .
	\end{eqnarray}
	Just as the coordinates are condensed, so are derivatives: 
	\begin{eqnarray}
		\frac{\partial}{\partial \mathbf{q}} = \left(\frac{\partial}{\partial q_1}, \dots, \frac{\partial}{\partial q_n} \right) \, , \qquad \frac{\partial}{\partial \bzeta} = \left(\frac{\partial}{\partial \zeta_1}, \dots, \frac{\partial}{\partial \zeta_n} \right) \, .
	\end{eqnarray}
	The same logic applies to differentials/measures. We will sometimes use derivatives and differentials to represent the canonical bases of the tangent and cotangent bundles of $\mathbb{R}^{2n}$ at the origin, which being isomorphic to $\mathbb{R}^{2n}$ can be just though of as $\mathbb{R}^{2n}$ itself. 
	\item The symbol ``$\cdot$'' represents an element-by-element product, \emph{e.g.}~$\mathbf{p} \cdot d\mathbf{q} =  p_1 \, dq_1 + \dots + p_n \, dq_n$.
	\item The wedge product is term-by-term, \emph{i.e.}~$d\mathbf{q} \wedge d\mathbf{p} =  dq_1 \wedge dp_1 + \dots + dq_n \wedge dp_n$.
	\item We fix $\hbar=1$ throughout the whole manuscript, and time is always a real parameter. The hamiltonian functions are always time-independent.
\end{itemize}

\subsection{The embedding $\mathbb{R}^{2n} \hookrightarrow \mathbb{C}^{2n}$}\label{subsec:emb}

The fact that $\mathbb{C}^{n} \sim \mathbb{R}^n \oplus i \mathbb{R}^n$ renders the employment of isomorphisms such as $\mathbf{z} \propto \mathbf{p} + i \mathbf{q}$ quite usual. However, as is well-known in the mathematical literature, it is often more enlightening to embed $\mathbb{R}^{2n}$ into $\mathbb{C}^{2n}$ when dealing with symplectic geometry \cite{FollandBook,NazaiBook}, as we will now review. We start by defining the \emph{complexification} as the map
\begin{eqnarray}
	\mathcal{W} : \mathbb{R}^{2n} \,\,\, &\longrightarrow \quad \! \mathbb{C}^{2n} \notag \\
	\quad (\mathbf{q},\mathbf{p}) &\longmapsto (\bzeta, \bzeta^*) = \mathcal{W} (\mathbf{q},\mathbf{p})
	\, , \quad \mathcal{W}  	
	= \frac{1}{\sqrt{2}}
	\begin{pmatrix}
	iI & I \\
	-iI & I 
	\end{pmatrix} \, , \label{comp}
\end{eqnarray}
with its inverse, the \emph{de-complexification}, given by
\begin{eqnarray}
	\!\! \mathcal{W}^{-1} : \mathbb{C}^{2n}  &\longrightarrow \quad \! \mathbb{R}^{2n} \notag \\
	\quad \,\,\,\, (\bzeta, \bzeta^*) &\longmapsto (\mathbf{q},\mathbf{p}) = \mathcal{W}^{-1} (\mathbf{q},\mathbf{p})
	\, , \quad \mathcal{W}  	
	= \frac{1}{\sqrt{2}}
	\begin{pmatrix}
	-iI & iI \\
	I & I 
	\end{pmatrix} \, . \label{decomp}
\end{eqnarray}

To see why the embedding given by \eqref{comp} is well suited from the symplectic point of view, we recall that a linear operator $\mathcal{S}$ fulfilling
\begin{equation}
	\mathcal{S}^T \mathcal{J} \mathcal{S} = \mathcal{J} \, , \quad \det \mathcal{S} = 1 \,, \quad \mathcal{J} =
	\begin{pmatrix}
	0 & I \\
	-I & 0 
	\end{pmatrix}  \, ,
\end{equation}
is called a \emph{symplectic matrix}, \emph{linear canonical transformation} or \emph{linear symplectomorphism}, and the set formed by all such operators composes the so called \emph{sympletic group}, usually denoted by $\text{Sp}(n)$. The complexification $\mathcal{W}$, however, obeys the slightly different relation
\begin{equation}
\mathcal{W}^T \mathcal{J} \mathcal{W} = i \mathcal{J} \,, \label{symp}
\end{equation} 
such that we can consider it to be \emph{$\lambda$-symplectic}, \emph{i.e}~a linear symplectomorphism with multiplier $\lambda=i$. In this way, the complexification of any linear operator $T$ on $\mathbb{R}^{2n}$ acts on $\mathbb{C}^{2n}$ through the similarity transformation
\begin{eqnarray}
	_\mathbb{C} : \text{Gl}(\mathbb{R}^{2n}) &\longrightarrow \text{Gl}(\mathbb{C}^{2n}) \notag \\
	  \,\, \qquad T &\longmapsto \quad \, T_\mathbb{C} = \mathcal{W} \, T \, \mathcal{W}^{-1} \, .
\end{eqnarray}
Evidently, the complexification can also be pushed-forward to act on general differential forms on $\mathbb{R}^{2n}$. In particular, it preserves the canonical form $\omega = d \mathbf{q} \wedge d\mathbf{p}$, but again with a multiplier $\lambda=i$: 
\begin{equation}
\omega = d\mathbf{q} \wedge d \mathbf{p} \quad \Longrightarrow \quad \omega_\mathbb{C} =  i \, d\bzeta^*\! \wedge d\bzeta \, , \label{forms}
\end{equation}
as can be verified by direct substitution using \eqref{comp} or \eqref{decomp}. 

\subsection{Complexified hamiltonian fields and dynamics}\label{subsec:dyn}

We now describe how the equations of motion change under the complexification embedding defined earlier. From \eqref{comp}, it is clear that the canonical basis vectors change to
\begin{equation}
	\frac{\partial}{\partial \mathbf{p}} = \frac{1}{\sqrt{2}} \left( \frac{\partial}{\partial \bzeta} + \frac{\partial}{\partial \bzeta^*}\right) \, , \quad \frac{\partial}{\partial \mathbf{q}} = \frac{i}{\sqrt{2}} \left( \frac{\partial}{\partial \bzeta} - \frac{\partial}{\partial \bzeta^*}\right) \, ,
\end{equation}
with inverse
\begin{equation}
	\frac{\partial}{\partial \bzeta} = \frac{1}{\sqrt{2}} \left( \frac{\partial}{\partial \mathbf{p}}  - \frac{i \partial}{\partial \mathbf{q}}\right) \, , \quad \frac{\partial}{\partial \bzeta^*} = \frac{1}{\sqrt{2}} \left( \frac{\partial}{\partial \mathbf{p}} + \frac{i \partial}{\partial \mathbf{q}}\right) \, .
\end{equation}
Then, either by direct substitution or using the pushforward of $\mathcal{W}$, we see that the gradient of a test function $f$ is mapped to
\begin{equation}
	\frac{\partial f}{\partial \mathbf{q}} \cdot \frac{\partial}{\partial \mathbf{q}} + \frac{\partial f}{\partial \mathbf{p}} \cdot \frac{\partial}{\partial \mathbf{p}} = \nabla f \quad \stackrel{\mathbb{C}}{\longmapsto} \quad \nabla_\mathbb{C} f_\mathbb{C} = \frac{\partial f_\mathbb{C}}{\partial \bzeta^*} \cdot \frac{\partial}{\partial \bzeta} + \frac{\partial f_\mathbb{C}}{\partial \bzeta} \cdot \frac{\partial}{\partial \bzeta^*} \, . \label{nablac}
\end{equation}
Notice the inversion taking place in the complexified gradient. We then see that Hamilton's equations transform as
\begin{equation}
	(\dot{\mathbf{q}},\dot{\mathbf{p}}) = \mathcal{J} \nabla H(\mathbf{q},\mathbf{p}) \quad \stackrel{\mathbb{C}}{\longmapsto} \quad (\dot{\bzeta}, \dot{\bzeta^*}) = \mathcal{J}_\mathbb{C} \nabla_\mathbb{C} H_\mathbb{C}(\bzeta, \bzeta^*) \, , \label{compham}
\end{equation}
where the complexified canonical matrix is symmetric and given by
\begin{equation}
	\mathcal{J}_\mathbb{C} = \mathcal{W} \mathcal{J} \mathcal{W}^{-1} = i
	\begin{pmatrix}
	I & 0 \\
	0 & -I 
	\end{pmatrix} \, . \label{Jc}
\end{equation}
Since $(\mathbb{R}^{2n},\omega)$ is identified here as the phase space, we will refer to $(\mathbb{C}^{2n},\omega_\mathbb{C})$ as the \emph{complexified} phase space. 

We now make a fundamental distinction between what the words ``complex'' and ``complexified'' refer to in this manuscript. There are two ways of interpreting the mapping in \eqref{compham}: The first is to consider it going left-to-right (as written), with real variables just \emph{complexified} and no information gained or lost in working with complex dynamics; And the second is to take complex dynamics as more fundamental, such that the reversed right-to-left map becomes a projection into real coordinates. If one considers a purely \emph{complex} phase space, with $\bzeta$ and $\bzeta^*$ independent of each other, the inverse mapping from complex to real will not be bijetive: Complex dynamics \emph{is richer} and involves phenomena impossible to achieve in real phase space. This is due to the existence of trajectories projecting to the same real ones, but with different imaginary parts, allowing for the semiclassical treatment of quantum processes forbidden in the real case, a prominent example being deep tunneling \cite{Aguiar2007,Heller1977,Klauder1994,Ozorio2010}. Thus, in a way, purely complex phase spaces do not really model classical mechanics, but effectively extend it (this extension can be identified with complex times \cite{Ozorio2010}). They also extend quantum mechanics, since complex positions and momenta do not necessarily fulfill the Poisson bracket identity $\{\mathbf{q},\mathbf{p}\}=I$, such that canonical quantization (and others) is not obvious and non-hermitian operators might be required \cite{Graefe2012}. In particular, the Heisenberg group, which underlies both classical and quantum mechanics, is extended. Thus, whenever we say ``complexification'', it must be understood that we are not referring to this general scenario of pure complex phase spaces, only to a mere parametrization in terms of complex variables -- although we will soon see its consequences are rather profound.

\subsection{Generating functions on complexified phase spaces}\label{subsec:gen}

We begin with a simple observation regarding an often ignored fact in the literature: There is no generating function that can be written as $S(\mathbf{q},\mathbf{p})$. If a function is responsible for implicitly defining a symplectomorphism, \emph{i.e.}~a canonical transformation, then it cannot have its domain fixed on initial variables -- it must also include the final ones. This is clearly expressed in the well-known generating functions of Goldstein \cite{GoldsteinBook}, which have as their domains four different position-momentum pairs: $(\mathbf{q}',\mathbf{q})$, $(\mathbf{q}',\mathbf{p})$, $(\mathbf{p}',\mathbf{q})$ and $(\mathbf{p}',\mathbf{p})$, where primed variables are final and non-primed, initial. The generating function usually denoted by $S(\mathbf{q},\mathbf{p};t)$ is actually the extended position generating funtion given by
\begin{equation}
	S(\mathbf{q}',\mathbf{q};t) = \int_\mathbf{q}^{\mathbf{q}'} \mathbf{p} \cdot d\mathbf{Q} - \int_0^t d\tau H(\mathbf{q},\mathbf{p}) \, , \label{genqq}
\end{equation}   
where the first integral is along the path joining $\mathbf{q}$ at $\tau=0$ to $\mathbf{q}'$ at $\tau=t$ by the hamiltonian flow (that is, by the solution to Hamilton's equations). Naturally, we can also write the function above as
\begin{equation}
	S(t) = \int_0^t d\tau \left[ \mathbf{p} \cdot \dot{\mathbf{q}} - H(\mathbf{q},\mathbf{p}) \right] \, , \label{genqt}
\end{equation}   
since $d\mathbf{Q}$ is a function of time. Expressions such as \eqref{genqq}, however, will be preferred whenever there is inherent interest in the variables with respect to which the generating function can be differentiated, \emph{e.g.}~for \eqref{genqq} we can use derivatives with respect to positions to obtain momenta \cite{ArnoldBook}. The numerical value of \eqref{genqq} and \eqref{genqt} is, of course, the same. 

The 1-form defined by the exterior derivative of \eqref{genqq}, namely
\begin{equation}
	\widetilde{\alpha} =  \mathbf{p} \cdot d\mathbf{q} - \mathbf{p}' \cdot d\mathbf{q}' - H(\mathbf{q},\mathbf{p}) \, dt \, , \qquad \widetilde{\alpha} = -dS(\mathbf{q}',\mathbf{q};t) \, ,
\end{equation}
just as the generating function itself, does not treat positions and momenta equally. This 1-form is a \emph{tautological form} on the extended product manifold $\mathbb{R}^{4n} \times \mathbb{R} = \{\mathbf{q}',\mathbf{p}',\mathbf{q},\mathbf{p};t\}$ \cite{AnaBook,SpivakBook}. It is a primitive for the extended canonical form 
\begin{equation}
\widetilde{\omega} = d\mathbf{q} \wedge d\mathbf{p} - d\mathbf{q}' \wedge d\mathbf{p}' + dH \wedge dt \, , \label{ext}
\end{equation}
that is, $d\widetilde{\alpha} = -\widetilde{\omega}$. Virtually all of classical mechanics is encoded in the relations between generating functions, tautological and canonical forms, and particularly important to us is the fact that tautological forms have an infinite number of primitives, related through Legendre transforms. This abundance allows for classical mechanics to be expressed in terms of a multitude of coordinates, and in semiclassical mechanics is directly responsible for us being able to express wave functions using positions, momenta, and others. For instance, the 1-form $ \mathbf{q} \cdot d\mathbf{p} - \mathbf{q}' \cdot d\mathbf{p}' - H(\mathbf{q},\mathbf{p}) \, dt$ will give birth \footnote{Generating functions can only be defined on the kernel of $\widetilde{\omega}$, \emph{i.e.}~the \emph{lagrangian} submanifolds $X$ for which $\widetilde{\omega}|_X=0$. To see this, note $\widetilde{\omega}|_X = 0 \Longrightarrow d\widetilde{\alpha} |_X = 0 \Longrightarrow \widetilde{\alpha} |_X = dS$. The graphs of symplectomorphisms are all lagrangian submanifolds w.r.t.~the extended canonical form \cite{AnaBook}. Since the hamiltonian flow is a family of symplectomorphisms w.r.t.~time, the generating functions used in this manuscript can always be defined.} to a generating function $S(\mathbf{p}',\mathbf{p};t)$ that involves momenta, not positions. This momentum generating function will describe dynamics just as its position equivalent, since they both obey the Hamilton-Jacobi equation. The unequal treatment of position and momentum used to obtain $S(\mathbf{q}',\mathbf{q};t)$ and $S(\mathbf{p}',\mathbf{p};t)$, however, will result in evolution being described using position and momentum representations, but the symmetrized 1-form
\begin{equation}
\widetilde{\alpha}_\text{W} = \left( \frac{\mathbf{p} \cdot d\mathbf{q} - \mathbf{q} \cdot d\mathbf{p}}{2} \right) + \left( \frac{\mathbf{q}' \cdot d\mathbf{p}' - \mathbf{p}' \cdot d\mathbf{q}'}{2} \right)  - H(\mathbf{q},\mathbf{p}) \, dt \, , \label{sw}
\end{equation}
while still a primitive for $\widetilde{\omega}$, places momentum and position in an equal footing and describes the evolution using the \emph{Segal-Bargmann representation}, to be discussed later on \cite{HallBook}. 

The label chosen for the 1-form in \eqref{sw} reflects its connection to Weyl (or symmetric \cite{Fierro2006,Aguiar2005}) quantization. Its most symmetric form in complexified coordinates is 
\begin{align}
\widetilde{\alpha}_{\text{W}, \mathbb{C}} &= i \left[ \left( \frac{\bzeta^* \cdot d\bzeta - \bzeta \cdot d\bzeta^*}{2} \right) - \left( \frac{\bzeta'^* \cdot d\bzeta' - \bzeta' \cdot d\bzeta'^*}{2} \right) \right] - H_\mathbb{C}(\bzeta,\bzeta^*) \, dt  \, , \label{alpha}
\end{align}
which is a primitive of the complexification \eqref{ext}, namely
\begin{equation}
\widetilde{\omega}_\mathbb{C} = i\, d\bzeta^* \wedge d\bzeta - i\, d\bzeta'^* \wedge d\bzeta' + dH_\mathbb{C} \wedge dt \, .
\end{equation}
The 1-form in \eqref{alpha} is a function of twice as many variables are needed, since half of them are dummy and can be obtained by complex conjugation. We can then choose any pair of initial and final variables and simplify the expression above using complexified Legendre transforms. The pair $(\bzeta^*, \bzeta')$ was favored by Weissman, which was the first to study these generating functions in physics \cite{Weissman1982}. Although this choice offers no particular advantage, it is unarguably favored by both physical and mathematical literatures \cite{Littlejohn1986,FollandBook}, and by adopting it we can compare our calculations with earlier works more easily. We then employ the substitutions $ \bzeta^* \cdot d\bzeta = d(|\bzeta|^2) - \bzeta \cdot d\bzeta^*$ and $\bzeta' \cdot d\bzeta'^* = d(|\bzeta'|^2) - \bzeta'^* \cdot d\bzeta'$ to isolate the differentials in \eqref{alpha} as functions of $(\bzeta^*, \bzeta')$, resulting in
\begin{align}
\widetilde{\alpha}_{\text{W},\mathbb{C}} &= i \left\{ d \left( \frac{ \vert \bzeta' \vert^2 + \vert \bzeta \vert^2 }{2}\right) - \left[ \bzeta \cdot d\bzeta^* + \bzeta'^* \cdot d \bzeta'  \right] \right\} \, . \label{alpha2}
\end{align}
The generating function is then given by 
\begin{align}
S_{\text{W},\mathbb{C}}(\bzeta',\bzeta^*;t) = i \left( \frac{ \vert \bzeta' \vert^2 + \vert \bzeta \vert^2}{2} \right) + F_{\text{W},\mathbb{C}}(\bzeta^*,\bzeta';t) \, , \label{gen2} 
\end{align}
where we have defined
\begin{align}
F_{\text{W},\mathbb{C}}(\bzeta',\bzeta^*;t) = - i \int \left( \bzeta \cdot d\bzeta^* + \bzeta'^* \cdot d\bzeta' \right) - \int_0^t d\tau H_\mathbb{C}(\bzeta,\bzeta^*) \, . \label{genz}
\end{align}
The reason for highlighting this function is due to it fulfilling
\begin{equation}
\frac{\partial F_{\text{W},\mathbb{C}}(\bzeta',\bzeta^*;t)}{\partial \bzeta^*} = -i \bzeta  \, ; \quad \frac{\partial F_{\text{W},\mathbb{C}}(\bzeta',\bzeta^*;t)}{\partial \bzeta'} = -i \bzeta'^* \, , \quad \frac{\partial F_{\text{W},\mathbb{C}}(\bzeta',\bzeta^*;t)}{\partial t} + H_\mathbb{C}(\bzeta,\bzeta^*) = 0 \, , \label{condC}
\end{equation}
such that, in agreement with \cite{Weissman1982}, $F_{\text{W},\mathbb{C}}(\bzeta',\bzeta^*;t)$ is the generating function of the complexified evolution from $(\bzeta,\bzeta^*)$ to $(\bzeta',\bzeta'^*)$, written in terms of the pair $(\bzeta^*, \bzeta')$.

The obscure expression in \eqref{genz} can be brought to a simpler form by changing the pair to $(\bzeta,\bzeta')$ using $\bzeta^* \cdot d\bzeta = d(|\bzeta|^2) - \bzeta \cdot d\bzeta^*$, resulting in
\begin{align}
F_{\text{W},\mathbb{C}}(\bzeta,\bzeta';t) = i \int_{\bzeta}^{\bzeta'} \bzeta^* \cdot d\mathbf{Z} - \int_0^t d\tau H_\mathbb{C}(\bzeta,\bzeta^*) - i |\bzeta|^2 \, , \label{gennotpop}
\end{align} 
which is sometimes found in literature \cite{Weissman1982,Weissman1983}. The simplest expression for the complexified generating function, however, arises by writing either the above or \eqref{gen2} without isolating absolute values, such that 
\begin{align}
S_{\text{W}, \mathbb{C}} (t) &= \int_0^t d\tau \left[ \frac{i}{2} \left( \bzeta \cdot \dot{\bzeta^*} - \bzeta^* \cdot \dot{\bzeta} \right) - H_\mathbb{C}(\bzeta,\bzeta^*) \right] \, . \label{genpop}
\end{align}
This function is clearly the most immediate primitive to $\eqref{alpha}$ and is by far the most popular in literature (\emph{e.g.}~\cite{Klauder1994,Aguiar2005,Klauder1979}). Although simpler and numerically identical to \eqref{gen2}, it makes it harder to draw the connections we shall develop in the following sections. 

\section{Linear theory}\label{sec:lin}

The simplest example of symplectomorphism is found in a linear setting, with mappings given by $\mathcal{M} \in \text{Sp}(n)$. Their quantization results in the group of unitary operators known as the \emph{metaplectic group}. If time-dependence is allowed, $\mathcal{M}(t)$ forms a 1-parameter family of linear symplectomorphisms, \emph{i.e.}~the path $\mathcal{M}(t)$ belongs to $\text{Sp}(n)$ for all $t \in \mathbb{R}$. In order to carry out quantization, however, it is necessary to choose a representation, and we shall now show not all representations are built equal.

\subsection{Linear symplectomorphisms}\label{subsec:lin}

Consider linear symplectomorphisms of the form
\begin{equation}
	(\mathbf{q}',\mathbf{p}') = \mathcal{M} (\mathbf{q},\mathbf{p}) \, , \quad \mathcal{M} =
	\begin{pmatrix}
		A & B \\
		C & D 
	\end{pmatrix} \, , \quad A,\,B,\,C,\,D \in \mathbb{R}^{n^2} \, , \label{linqp}
\end{equation}
with positions and momenta necessarily real. The complex equivalent of the above system is given by employing the similarity transformation $\mathcal{M}_\mathbb{C} = \mathcal{W} \mathcal{M} \mathcal{W}^{-1}$:
\begin{equation}
	(\bzeta', \bzeta'^*) = \mathcal{M}_\mathbb{C} (\bzeta, \bzeta^*) \, , \quad \mathcal{M}_\mathbb{C} = 
	\begin{pmatrix}
	\Lambda & \Gamma  \\
	\Gamma^* & \Lambda^*
	\end{pmatrix} \, , \label{linzz}
\end{equation}
with
\begin{equation}
	\Lambda = \frac{1}{2} \left[ \left( D + A\right) + i \left( C - B \right) \right] \, , \quad \Gamma = \frac{1}{2} \left[ \left( D - A \right) - i \left( C + B \right) \right] \, . \label{compzz} 	
\end{equation}

The time-independent position generating function $S(\mathbf{q}',\mathbf{q})$ is obtained from \eqref{linqp} by writing $\mathbf{p}'$ and $\mathbf{p}$ as exclusive functions of $(\mathbf{q},\mathbf{q}')$, then finding a quadratic form fulfilling the differential conditions required, namely 
\begin{equation}
	\frac{\partial S(\mathbf{q}',\mathbf{q})}{\partial \mathbf{q}} = - \mathbf{p} \, ; \qquad \frac{\partial S(\mathbf{q}',\mathbf{q})}{\partial \mathbf{q}'} = \mathbf{p}' \, , \label{mom}
\end{equation} 
the well-known answer being 
\begin{equation}
	S(\mathbf{q}',\mathbf{q}) = \frac{1}{2} \left[ \mathbf{q}' \cdot (B^{-1} A) \mathbf{q}' + \mathbf{q} \cdot (D  B^{-1}) \mathbf{q} - 2 \mathbf{q}' \cdot (B^{-1}) \mathbf{q} \right] \, .	\label{genclass}
\end{equation}
It is clear that the generating function above can only be defined for non-singular $B$, in which case we say the matrix $\mathcal{M}$ is \emph{free} \cite{deGossonBook}. If, on the other hand, one looks for a complex generating function $S_{\text{W},\mathbb{C}}(\bzeta',\bzeta^*)$ by writing $\bzeta$ and $\bzeta'^*$ in terms of $(\bzeta^*,\bzeta')$ in \eqref{linzz} and integrating the first two differential conditions in \eqref{condC}, the result is \footnote{Note this generating function has the exact same form as \eqref{gen2}.} 
\begin{equation}
S_{\text{W},\mathbb{C}}(\bzeta^*,\bzeta') = \frac{i}{2} \left( |\bzeta^*|^2 + |\bzeta'|^2 \right) - \frac{i}{2} \left[ \bzeta' \cdot \left( \Gamma^* \Lambda^{-1} \right) \bzeta' - \bzeta^* \cdot \left( \Lambda^{-1} \Gamma \right) \bzeta^* + 2 \bzeta^* \cdot (\Lambda^{-1}) \bzeta' \right] \, .	\label{gencomp}
\end{equation}
In \ref{App:A} we show that the matrix $\Lambda$ in the equation above, defined in \eqref{compzz}, is \emph{always} non-singular, as long as $\mathcal{M}$ is symplectic. Thus, while a description in terms of the position generating function is limited to free symplectic matrices, the complexified generating function above works for any element of $\text{Sp}(n)$.

The contrast between employing either \eqref{genclass} or \eqref{gencomp} becomes more stringent when we allow the symplectic matrix in \eqref{linqp} to be time-dependent. In this case, the 1-parameter family $\mathcal{M}(t)$ belongs to $\text{Sp}(n)$ for all $t \in \mathbb{R}$, a situation found \emph{e.g.}~in the case of hamiltonian flows obtained from quadratic hamiltonian functions. The allowance of time-dependence means that we can start with a free $\mathcal{M}(0)$ and, as time grows, hit a bump at which $B(t)$ becomes singular. When this happens, evolution can no longer be described using the time-dependent position generating function $S(\mathbf{q}',\mathbf{q};t)$, and is known as a \emph{caustic} in position space. Notice this never happens for $S_{\text{W},\mathbb{C}}(\bzeta',\bzeta^*;t)$, which is able to describe the path traced by $\mathcal{M}(t)$ in $\text{Sp}(n)$ for all times. A description in terms of complex variables then naturally allows us to bypass caustics, while in real phase space we were somewhat stuck with the singularities in $S(\mathbf{q}',\mathbf{q};t)$. 

The usual procedure to evade the caustics in $S(\mathbf{q}',\mathbf{q};t)$ is to move to another coordinate system, that is, to describe \eqref{linqp} using one of the conjugate generating functions $S(\mathbf{p}',\mathbf{p};t)$, $S(\mathbf{p}',\mathbf{q};t)$ or $S(\mathbf{q}',\mathbf{p};t)$. For each of these a different component of $\mathcal{M}(t)$ appears inversed, \emph{i.e.}~$C(t)$ for $S(\mathbf{p}',\mathbf{p};t)$ or $A(t)$ for $S(\mathbf{q}',\mathbf{p};t)$. It is easy to see these generating functions cannot be simultaneously singular: If this happened then $\mathcal{M}(t)$ would itself be singular, contradicting its symplecticity. However, these conjugate functions will also develop caustics themselves, such that we end up being forced to go back and forth between at least two of them in order to describe the path traced by $\mathcal{M}(t)$. The complex generating function in \eqref{gencomp}, however, is able to describe the whole evolution on its own. 

Caustic avoidance is a first evidence that complex variables might present advantages over the usual position and momentum ones, but this is only true if the complex variables are obtained by complexification, \emph{i.e.}~$A+D$ and $C-D$ in \eqref{compzz} need to be real. If these are complex $\Lambda$ can become singular, such as for $A+B=2I$ and $C-B=2iI$, a choice incompatible with the symplecticity of $\mathcal{M}$ due to the imaginary determinant \footnote{As stated earlier, complex symplectic matrices need the underlying structure of classical mechanics to be modified, such that $\lambda$-symplectomorphisms take the place of the usual ones in order to deal with complex determinants.}. Thus, complex phase spaces \emph{do} have caustics, but complexifications of real phase spaces do not \cite{Littlejohn1986,FollandBook,deGossonBook}.

\subsection{Metaplectic families and their representations}\label{subsec:meta}

The quantization of $\text{Sp}(n)$ results in a unitary group known as the \emph{metaplectic group}, $\text{Mp}(n)$ \cite{FollandBook,deGossonBook,OzorioBook,GutzwillerBook}. It forms a double cover of $\text{Sp}(n)$, such that each symplectic matrix has two unitary operators associated to it in $\text{Mp}(n)$. The position representation of the metaplectic family quantizing $\mathcal{M}(t)$ is given by
\begin{eqnarray}
	\!\!\!\!\!\!\!\!\!\!\!\!\!\!\!\!\!\!\!\!\!\!\!\! \langle \mathbf{q}' | \widehat{\mathcal{M}}(t) | \mathbf{q} \rangle &= \sigma \,(2\pi)^{-\frac{n}{2}} \left\vert \det B(t) \right\vert^{-\frac{1}{2}} \exp \left\{ \frac{i}{2} \bigg( \mathbf{q}' \cdot \left[ B^{-1}(t) \, A(t) \right] \mathbf{q}' \right. \notag \\
	&\qquad \qquad \left. + \mathbf{q} \cdot \left[ D(t) \, B^{-1}(t) \right] \mathbf{q} - 2 \mathbf{q}' \left[ B^{-1}(t) \right] \mathbf{q} \bigg) - \frac{i \pi \mu}{2} \right\} \, , \label{Uqq}
\end{eqnarray}
where the index $\sigma = \pm 1$ indicates the two possible quantizations of each $\mathcal{M}(t)$ for a fixed time. It is clear that $\widehat{\mathcal{M}}(t)$ cannot be expressed in the position representation if $\mathcal{M}(t)$ is not free, just as in this case its classical path cannot be described using position generating functions. We are then forced to switch to a non-singular conjugate generating function, which quantum mechanically amounts to changing representation. The intertwining between generating functions across caustics, however, is responsible for an accumulated phase known as the \emph{Maslov index}, $\mu$. Written as in \eqref{Uqq}, this index is nothing but the number of caustics encountered alongside the trajectory linking $(\mathbf{q},\mathbf{p})$ and $(\mathbf{q}',\mathbf{p}')$ \cite{Littlejohn1992,ArnoldBook,deGossonBook,OzorioBook,GutzwillerBook,Ozorio2014}.

The map taking a symplectic matrix to its corresponding metaplectic operators is exact, such that the classical and quantum objects are exactly the same. Thus, the caustics appearing in \eqref{Uqq} are not at all failures, but fundamental components inherent to quantum dynamics. As an example we can take the general caustic at $t=0$, with $\hat{\mathcal{M}}(0) = \hat{I}$. Here, the position representation of the metaplectic operator is just $\delta(\mathbf{q}' - \mathbf{q})$, and is exactly what the caustic is reproducing: A divergence in the quantum propagator. There is nothing wrong with a blowup in \eqref{Uqq}, just as there is nothing wrong with $\delta(\mathbf{q}' - \mathbf{q})$, which is a perfectly well-behaved distribution \cite{Littlejohn1986,SchwartzBook}. If, however, we had chosen a representation in which the basis elements do not contract to Dirac deltas, there would be no divergences to be reproduced by their expression in terms of classical generating functions. In particular, the coherent state basis builds the \emph{Segal-Bargmann} (S-B) representation, and since
\begin{equation}
	\langle \bzeta^* | \bzeta' \rangle = \exp \left( \frac{-|\bzeta^*|^2-|\bzeta'|^2}{2} + \bzeta \cdot \bzeta' \right) \neq \delta(\bzeta' - \bzeta^*) \, ,
\end{equation}
it provides a normalizable description that remains in Hilbert space. In this representation, the family $\widehat{\mathcal{M}}(t)$ is expressed using the complex generating function \eqref{gencomp} as \cite{Littlejohn1986,FollandBook}
\begin{eqnarray}
	\!\!\!\!\!\!\!\!\!\!\!\!\!\!\!\!\!\!\!\!\!\!\!\! \langle \bzeta' | \widehat{\mathcal{M}}(t) | \bzeta^* \rangle &= \left[ \frac{\exp \left[ - \frac{1}{2} \left( |\bzeta^*|^2 + |\bzeta'|^2 \right) \right]}{\sqrt{ \det \Lambda(t)}} \right] \exp \bigg\{ \frac{1}{2} \bigg( \bzeta' \cdot \left[ \Gamma^*(t) \, \Lambda^{-1}(t) \right] \bzeta'  \notag \\[8pt]
	&\qquad \qquad  - \bzeta^* \cdot \left[ \Lambda^{-1}(t) \, \Gamma(t) \right] \bzeta^* + 2 \bzeta^* \cdot \left[ \Lambda^{-1}(t) \right] \bzeta' \bigg) \bigg\} \, . \label{Uzz}
\end{eqnarray}
The discontinuous phase jumps that happen in \eqref{Uqq} across a caustic end up being mapped into branch changes \footnote{Branch changes and caustics do not happen at the same spacial and/or temporal places. See Sec.~\ref{subsec:pre}.} of the complex amplitude in \eqref{Uzz}, \emph{i.e.}~by tracking the continuity in the complex pre-factor we are directly establishing the correct phase changes in the complex propagator \cite{Kay2005}. The choice of metaplectic leaf, specified in \eqref{Uqq} by $\sigma=\pm 1$, is included in \eqref{Uzz} as an indeterminacy in its initial square root \footnote{Absolute values of jacobian determinants are absent in the S-B representation because the real orientation is preserved by any complexified map. Proofs for $\mathbb{C}$-linear and holomorphic maps can be found in \cite{VolkerBook}. The proof for general maps uses Sylvester's theorem and is discussed in \emph{e.g.}~\cite{CrossBook} and \cite{FarautBook}.} sign \cite{FollandBook}. Note that the ``over-completeness of the coherent state basis'', often seen as an undesirable aspect of the S-B representation, is the fundamental reason for its absence of caustics. 

The mapping taking \eqref{Uqq} to \eqref{Uzz} is a composition of two \emph{S-B transforms} \cite{FollandBook,HallBook,Bargmann1961}. More important to us is the inverse transform, given in our notation by
\begin{equation}
	\langle \mathbf{x} | \bphi \rangle = \int d\bzeta^* \langle \mathbf{x} | \bzeta^* \rangle \langle \bzeta^* | \bphi \rangle \, ,
\end{equation}
where $d\bzeta^*$ is the Lebesgue measure on $\mathbb{C}^n$. This transform maps the state $\langle \bzeta^* | \bphi \rangle$ in the S-B representation back to the position one, and its integral kernel is given by what is known in physics as a \emph{Schr\"odinger coherent state}, which is unnormalized \cite{FollandBook,GazeauBook,SweitseBook}. The lack of normalization is fundamental in order to contain the gaussian measure $\mu(d\bzeta^*) = \exp(-|\bzeta^*|^2) \, d\bzeta^*$, with which the S-B representation is equipped. Thus, the map taking \eqref{Uzz} to \eqref{Uqq} is given by the composition
\begin{align}
\langle \mathbf{x}' | \widehat{\mathcal{M}}(t) | \mathbf{x} \rangle = N \int d\bzeta^* d\bzeta' \langle \mathbf{x}' | \bzeta' \rangle \langle \bzeta' | \widehat{\mathcal{M}}(t) | \bzeta^* \rangle \langle \bzeta^* | \mathbf{x} \rangle \, , \label{sb}
\end{align}
where $N$ is a normalization factor. The Schr\"odinger coherent states in the above equation are particularly simple when expressed in terms of complex coordinates, being given by
\begin{align}
\begin{cases}
\langle \bzeta^* \vert \mathbf{x} \rangle = \pi^{-\frac{n}{4}} \exp \left[ \dfrac{1}{2} \left( - \mathbf{x} \cdot \mathbf{x} - 2 i \sqrt{2} \, \bzeta^* \cdot \mathbf{x} + \bzeta^* \cdot \bzeta^* - \vert \bzeta^* \vert^2 \right) \right] \\[8pt]
\langle \mathbf{x}' \vert \bzeta' \rangle = \pi^{-\frac{n}{4}} \exp \left[ \dfrac{1}{2} \left( - \mathbf{x}' \cdot \mathbf{x}' + 2 i \sqrt{2} \, \bzeta' \cdot \mathbf{x}' + \bzeta' \cdot \bzeta' -\vert \bzeta' \vert^2  \right) \right]
\end{cases} \label{cohs} \, ,
\end{align}
unlike \emph{e.g.}~Klauder coherent states, which have more complicated expressions. Naturally, by performing the gaussian integrals arising from substituting \eqref{Uzz} into \eqref{sb}, the expression for \eqref{Uqq} is exactly recovered \cite{Littlejohn1986}. Note that the S-B transform and its inverse end up complexifying and de-complexifying phase space, respectively, which is why we rely so much on the mappings \eqref{comp} and \eqref{decomp} \cite{Heller1987}	.

\section{Semiclassical approximations}\label{sec:semi}

We now extend what was presented in the earlier section to arbitrary hamiltonian flows, which are also 1-parameter families of [generally non-linear] symplectomorphisms. An exact extension is of course impossible, since it would imply that quantum and classical mechanics are identical, but an approximate link is established through the use of Stationary Phase Approximations. We then review the Initial Value Representation method and obtain an expression for the semiclassical propagator in position representation as a sequence of inverse Segal-Bargmann transforms of its complexified equivalent. This results in the Herman-Kluk propagator, a cornerstone of computational chemistry.

\subsection{Semiclassical propagators in the position and Segal-Bargmann representations}\label{subsec:sqb}

The early work by van~Vleck \cite{VanVleck1928}, together with the phase correction devised by Gutzwiller and others \cite{GutzwillerBook}, concentrated on generalizing \eqref{Uqq} from symplectic matrices to general hamiltonian flows, which are a particular type of 1-parameter family of non-linear symplectomorphisms. By writing the quantum propagator in position representation as
\begin{equation}
	K (\mathbf{x}',\mathbf{x};t) = \langle \mathbf{x}' | \widehat{U}(t) | \mathbf{x} \rangle \, , \qquad \widehat{U}(t) = e^{-i t \widehat{H}} \, ,
\end{equation}
the van~Vleck-Gutzwiller (vV-G) propagator approximates it as
\begin{equation}
	K_{\text{vV-G}}(\mathbf{x}',\mathbf{x};t) = \left( 2\pi i \right)^{-\frac{n}{2}} \sum_p \left\vert \det \left( \frac{\partial^2 S(\mathbf{x}',\mathbf{x};t)}{\partial \mathbf{x}' \, \partial \mathbf{x}} \right) \right\vert^{\frac{1}{2}}  \exp \left( i \left[ S(\mathbf{x}',\mathbf{x};t) - \frac{\pi \mu}{2} \right] \right) \, . \label{vV}
\end{equation}  
The generating function appearing above is the one in \eqref{genqq}, and the sum runs over all the trajectories that connect $\mathbf{x}$ at $\tau=0$ to $\mathbf{x}'$ at $\tau=t$ (thus, over all initial momenta fulfilling the first equation in \eqref{mom}). The determinant in the amplitude is now a component of the \emph{monodromy matrix} $\mathbb{M}$, which is symplectic and defined in terms of positions $\mathbf{x}$ and momenta $\mathbf{y}$ as
\begin{align}
	\mathbb{M}(\mathbf{x},\mathbf{y};t) =
	\begin{pmatrix}
	\dfrac{\partial \mathbf{x}'(\mathbf{x},\mathbf{y};t)}{\partial \mathbf{x}} & \dfrac{\partial \mathbf{x}'(\mathbf{x},\mathbf{y};t)}{\partial \mathbf{y}} \\[8pt]
	\dfrac{\partial \mathbf{y}'(\mathbf{x},\mathbf{y};t)}{\partial \mathbf{x}} & \dfrac{\partial \mathbf{y}'(\mathbf{x},\mathbf{y};t)}{\partial \mathbf{y}}	
	\end{pmatrix} \label{mon} \, ,
\end{align}
where $(\mathbf{x}',\mathbf{y}')$ represent the hamiltonian flow as a function of $(\mathbf{x},\mathbf{y})$

The linear setting in the earlier section resulted in the simple expressions \eqref{Uqq} and \eqref{Uzz} because, for linear systems, there is a single trajectory connecting two phase space points for a fixed time $t$ -- indeed, \eqref{vV} is brought to \eqref{Uqq} for linear flows. A further characteristic that renders the linear scenario particularly simple is the fact that, since the matrix in \eqref{linqp} cannot depend on phase space points, its caustics are exclusive functions of time. Besides, by the exactness of quantization in this case, there is no ``semiclassical failure'', since any divergence in \eqref{Uqq} arises due to the divergences in the quantum propagator itself.

The non-linear scenario is considerably more intricate. The correspondence between quantum and semiclassical propagators is not exact anymore, and caustics become functions of time \emph{and} space simultaneously. This allows for the semiclassical propagator being truly incapable of reproducing quantum evolution, its failures traceable back to classical mechanics. It is evident these failures are due to caustics, a problem worsened by the fact that even the caustic structure of simple non-quadratic hamiltonians can be excruciatingly complicated, as will be shown in Sec.~\ref{sec:kerr}.

A particularly special caustic it the one at $t=0$, since for this one the vV-G propagator's divergence is a correct one (the quantum propagator also diverges). This might lead to vV-G being reasonably good for extremely short times, when the quantum propagator is still close to a Dirac delta; followed by gradually degrading quality for an ``intermediate short-time regime'', where the quantum propagator is farther from a delta but the semiclassical one is still in the vicinity of the $t=0$ caustic; and finally regaining accuracy as we move away from the short-time regime. In Sec.~\ref{subsec:auto} we will see this is precisely what happens. 

If we are not at the time-origin, however, the quantum propagator can only become a Dirac delta again for some specific situations. One of these is the case of periodic systems, for which the evolution operator is the identity at each multiple of the system's period. A second one takes place for hamiltonians that are exclusive functions of the position operator, since $K(\mathbf{x}',\mathbf{x};t) = U(\mathbf{x};t) \, \delta(\mathbf{x}'-\mathbf{x})$. For a general hamiltonian that is a function of both momentum and position, however, quantum divergences cannot happen: The quantum propagator is smooth. Caustics are then seen as an exclusively classical problem, its roots traced to choosing representations in terms of non-normalizable bases, as described in Secs.~\ref{subsec:lin} and \ref{subsec:meta}.    

We now come back to the complexified case. The extension of \eqref{Uzz} to general hamiltonian flows was solidified in the early 1980s with the work of Weissman \cite{Weissman1982,Weissman1983}, based on a number of earlier developments (\emph{e.g.}~\cite{Heller1977,Miller1974}). His result, often referred to as the \emph{semiclassical coherent state propagator}, is nothing but the semiclassical propagator in the S-B representation, given by
\begin{equation}
	\langle \bzeta' | \widehat{U}(t) | \bzeta^* \rangle \approx \sum_{\bzeta} \left[ i \det \left( \frac{\partial^2 F_\mathbb{C}(\bzeta',\bzeta^*;t)}{\partial \bzeta^* \, \partial \bzeta'} \right) \right]^{\frac{1}{2}} \exp \left[ - \frac{1}{2} \left( |\bzeta^*|^2 + |\bzeta'|^2 \right) + i F_\mathbb{C}(\bzeta',\bzeta^*;t)  \right] \, . \label{zZ}
\end{equation} 
This was re-derived several times and shown to be identical to the semiclassical propagator obtained by Klauder through coherent state path integrals \cite{Weissman1983,Klauder1979}. In particular, it was shown that the expression above is only true if quantization using the Weyl ordering rule is assumed \cite{Aguiar2005}, which in our presentation is obvious due to its phase being exactly the generating function in \eqref{gen2}. If different orderings are used, a correction factor needs to be included in the phase, as discussed in great detail in \cite{Aguiar2005,Baranger2001}. As in Sec.~\ref{subsec:meta}, the phase jumps across branches are directly included in the complex pre-factor. The sum over trajectories now runs over the second complexified equation in \eqref{condC}, and has been carefully examined in several instances \cite{Grossmann1998,Aguiar2007,Klauder1994,Fierro2006,Aguiar2005,Kay2005,Baranger2001,Kay1994,Tannor2018}. In the complexified case it is equivalent to the real root search, but for complex phase spaces purely imaginary phenomena, such as Stokes divergences, generally occur \cite{Klauder1994}. 

To avoid referring to \eqref{zZ} as the ``semiclassical Weyl-ordered propagator in the S-B representation'', we rename it plainly as the \emph{S-B propagator}. Now, just as in the case of symplectic matrices, we expect \eqref{zZ} to be mapped to \eqref{vV} by two inverse S-B transforms. This correspondence, however, is not obtained exactly: Integral transforms are not endomorphisms in the ``category'' of semiclassical propagators. One needs to evaluate the transforms using Stationary Phase Approximations (SPAs), and only then can we exchange between different quantum representations/classical generating functions. Notwithstanding the fact that the semiclassical propagators \eqref{vV} and \eqref{zZ} are in correspondence through SPAs, we observe the same phenomenon described in Sec.~\ref{subsec:meta}: The vV-G propagator is plagued by divergences, whilst the S-B propagator is strictly continuous. This can be easily seen by using \eqref{condC} to rewrite the pre-factor in \eqref{zZ}:
\begin{equation}
	\left[ i \det \left( \frac{\partial^2 F_\mathbb{C}(\bzeta',\bzeta^*;t)}{\partial \bzeta^* \partial \bzeta'} \right) \right]^{\frac{1}{2}} = \left[ \det \left( \frac{\partial \bzeta'(\bzeta,\bzeta^*;t)}{\partial \bzeta} \right) \right]^{-\frac{1}{2}} = \big[ \det \Lambda(\bzeta^*,\bzeta;t) \big]^{-\frac{1}{2}} \, . \label{detCC}
\end{equation} \\

Before moving on we briefly stop to dedicate some attention to the complexified monodromy matrix appearing in the expression above. Despite the monodromy matrix not really fitting the context of linear systems as described in Sec.~\ref{sec:lin}, it can be connected to its complexification in an equivalent manner. For this, we remind the reader that the monodromy matrix has its dynamics defined by the ordinary differential equation 
\begin{equation}
	\dot{\mathbb{M}} = \mathcal{J} \, \text{Hess}(H) \, \mathbb{M} \, ,
\end{equation}
with $\text{Hess}$ representing the hessian \cite{NazaiBook}. This equation is linear, such that by similarity transformations we get
\begin{equation}
	\dot{\mathbb{M}}_\mathbb{C} = \mathcal{J}_\mathbb{C} \, \text{Hess}_\mathbb{C}(H_\mathbb{C}) \, \mathbb{M}_\mathbb{C} \, , \label{monc}
\end{equation}
with $\mathcal{J}_\mathbb{C}$ defined in \eqref{Jc} and the complex hessian obtainable \emph{e.g.}~from differentiating \eqref{nablac}. This shows the complexified monodromy has components exactly equal to the ones in \eqref{compzz}, except that now $A,\,B,\,C$ and $D$ are given by \eqref{mon}. 

\subsection{Initial value representation for the van~Vleck-Gutzwiller propagator}

Although the S-B propagator \eqref{zZ} is free from caustics, it suffers from a severe drawback also found in \eqref{vV}: The sum over classical trajectories. Except for a tiny number of systems, Hamilton's equations have to be solved numerically, and looking for trajectories entering the root-search is a computationally demanding task. After an insight by Miller \cite{Miller1970}, chemists started substituting the sum over trajectories by a full integral with respect to initial momenta, a procedure that can be mnemonically written as
\begin{equation}
	\int d\mathbf{q}' \sum_\mathbf{p} \left\vert \det \left( \frac{\partial \mathbf{q}'}{\partial \mathbf{p}} \right) \right\vert^{-\frac{1}{2}} \longmapsto \int d\mathbf{p} \left\vert \det \left( \frac{\partial \mathbf{q}'}{\partial \mathbf{p}} \right) \right\vert^{\frac{1}{2}}  = \int d\mathbf{p} \left\vert \det \left( \frac{\partial^2 S(\mathbf{q},\mathbf{q}';t)}{\partial \mathbf{q}' \, \partial \mathbf{q}} \right) \right\vert^{-\frac{1}{2}} \, , \label{ivrq}
\end{equation}
where in the last equation we used \eqref{mom}. Instead of looking for all possible momenta that define the trajectories entering the semiclassical sum, the substitution above simply sums over \emph{all} initial momenta and results in an \emph{Initial Value Representation} (IVR). The unfamiliar reader is directed to \cite{Miller2001} for a review and \cite{HellerBook} for a nice geometrical discussion on this substitution, for which we give a bit of mathematical context in \ref{App:B}. A welcome consequence of the pre-factor inversion in \eqref{ivrq} is that the previous divergences turn to converge toward zero as a caustic is approached. Another collateral effect is that, since the pre-factor acts as a weight, the contributions that were large for vV-G become small in the IVR. At this point it is not obvious whether or not this is desirable, a point we shall expand in Sec.~\ref{subsec:fail}.

Applying the substitution \eqref{ivrq} to \eqref{vV} results in a typical IVR expression for the vV-G propagator, given by
\begin{equation}
	K_\text{IVR}(\mathbf{x}',\mathbf{x};t) = \left( 2\pi i \right)^{-\frac{n}{2}} \int d\mathbf{q} \, d\mathbf{p} \,\left\vert \det \left( \frac{\partial \mathbf{q}'}{\partial \mathbf{p}} \right) \right\vert^{\frac{1}{2}}  \exp \left( i \left[ S(\mathbf{q},\mathbf{q}';t) - \frac{\pi \mu}{2} \right] \right) \delta \left( \mathbf{q}' - \mathbf{x}' \right) \, . \label{IVR}
\end{equation}  
The Dirac delta is a reminiscent of the root-search problem, and is in fact equivalent to it. This can be seen by noting that, by the compositional property of the delta (see \ref{App:B}), we have
\begin{equation}
\int d\mathbf{p} \, \delta \left[ \mathbf{q}'(\mathbf{x},\mathbf{p};t) - \mathbf{x}' \right] = \sum_{\ker \mathbf{q}'(\mathbf{x},\mathbf{p};t)} \left\vert \det \left( \frac{\partial \mathbf{q}'}{\partial \mathbf{p}} \right) \right\vert^{-1} \, ,
\end{equation}
where the kernel of $\mathbf{q}'$ is to be searched for w.r.t.~the initial momenta, which are the integration variables. 

An important characteristic of the IVR in \eqref{IVR} is that it cannot be immediately used to calculate wave functions, and one is forced to chose between limiting its use to numbers, \emph{e.g.}~$\langle \bpsi | \hat{U}(t) | \bphi \rangle$, or to develop some clever strategy to substitute the Dirac delta by something smoother \cite{Heller1991-2,Kay1993}. Several important quantities in chemistry and physics, however, are of the desired form for \eqref{IVR} to be promptly employed, a prominent example being the autocorrelation function
\begin{equation}
	C(t) = \langle \bpsi | \hat{U}(t) | \bpsi \rangle \, ,\label{auto}
\end{equation}
which is immediately seen to have the IVR expression
\begin{equation}
	C(t) \approx \left( 2\pi i \right)^{-\frac{n}{2}} \int d\mathbf{q} \, d\mathbf{p} \,\left\vert \det \left( \frac{\partial \mathbf{q}'}{\partial \mathbf{p}} \right) \right\vert^{\frac{1}{2}}  \exp \left( i \left[ S(\mathbf{q},\mathbf{q}';t) - \frac{\pi \mu}{2} \right] \right) \langle \bpsi^* \vert \mathbf{q} \rangle \langle \mathbf{q}' \vert  \bpsi \rangle \,  \label{autoIVR}
\end{equation}  
after the $\mathbf{x}'$ integrals are performed \cite{Miller2001}. 

Although the autocorrelation is an important quantity, one might be interested in the semiclassical approximation to more general objects. Several methods to obtain workable IVRs that could calculate proper wave functions were then developed, especially by Kay \cite{Kay1994}, who was one of the first to report problems with the slow convergence and errors due to the oscillatory behavior of several IVR expressions \cite{Kay1993}. Interestingly, the Wigner representation allows for IVRs capable of calculating evolved Wigner functions directly, the final expression being free of Dirac deltas \cite{Ozorio1998,Ozorio2013}. As IVR techniques have nowadays developed into a proper branch of computational chemistry, we will limit our exposition to an analysis of \eqref{IVR} and redirect the interested reader to the seminal papers \cite{Kay1993}, \cite{Miller2001} and \cite{Kay1994}.

\subsection{Initial value representation for the Segal-Bargmann propagator}

The absence of caustics in the S-B representation is a stark motivation to pursue an IVR using the S-B propagator \eqref{zZ}. We start by introducing it into \eqref{sb}, with $\widehat{\mathcal{M}}$ substituted by the evolution operator $\widehat{U}(t)$, to obtain an expression for the position element in terms of complexified variables, 
\begin{equation}
	K(\mathbf{x}',\mathbf{x};t) \approx N \int d\bzeta^* d\bzeta' \sum_{\bzeta} 	\left[ i \det \left( \frac{\partial^2 F_\mathbb{C}(\bzeta',\bzeta^*;t)}{\partial \bzeta^* \partial \bzeta'} \right) \right]^{\frac{1}{2}} e^{i S_{\text{W},\mathbb{C}}(t)} \langle \mathbf{x}' \vert \bzeta' \rangle \langle \bzeta^* | \mathbf{x} \rangle \, , \label{pre-ivr}
\end{equation}
where we choose the simplest expression in \eqref{genpop} for the generating function (equivalent to both \eqref{gen2} and \eqref{gennotpop}, being still a function of $\bzeta$ and $\bzeta^*$). To obtain the IVR, notice that the substitution \eqref{ivrq} is translated to the S-B representation as
\begin{equation}
	\int d\bzeta' \sum_\zeta \left[ \det \left( \frac{\partial \bzeta'}{\partial \bzeta} \right) \right]^{-\frac{1}{2}} \longmapsto \int d\bzeta \, \left[ \det \left( \frac{\partial \bzeta'}{\partial \bzeta} \right) \right]^\frac{1}{2} = \int d\bzeta \, \sqrt{\det \Lambda}  \, , \label{ivrz}
\end{equation}
such that all we need to do is to employ it in \eqref{pre-ivr} to get
\begin{equation}
	K(\mathbf{x}',\mathbf{x};t) \approx N \int d\bzeta^* d\bzeta \, \sqrt{\det \Lambda(\bzeta^*,\bzeta;t)} \, e^{i S_{\text{W},\mathbb{C}}(t)} \langle \mathbf{x}' \vert \bzeta'(\bzeta^*,\bzeta;t) \rangle \langle \bzeta^* | \mathbf{x} \rangle \, , \label{pre-preivr}
\end{equation}
with $N=\pi^{-n}$ obtained by requiring the propagator to be 1 for $t=0$. While the substitution in \eqref{ivrq} reverberates in multiple characteristics of the IVR due to the presence of caustics, their absence in the S-B representation means that \eqref{ivrz} does not really change anything except getting rid of the root-search.

The formula in \eqref{pre-preivr} can be seen as a result in itself: It is an IVR for the semiclassical propagator in position representation, obtained from the S-B propagator as a composition of two inverse S-B transforms. It will be free of caustics and phase jumps are implicitly included in the continuity of its complex pre-factor. However, by de-complexifying the IVR, we can write the propagator in terms of our real trajectories explicitly. To this end, we note the de-complexified Schr\"odinger coherent states in position representation are just
\begin{eqnarray}	
	\begin{cases}
	\,\,\, \langle \bzeta^*(\mathbf{q},\mathbf{p}) | \mathbf{x} \rangle \,\,= \pi^{-\frac{n}{4}} \exp \left[ - \dfrac{|\mathbf{x}-\mathbf{q}|^2}{2} - i \mathbf{p} \cdot \left( \mathbf{x}-\dfrac{\mathbf{q}}{2} \right) \right]  \\[8pt]
	\langle \mathbf{x}' \vert \bzeta'(\mathbf{q},\mathbf{p};t) \rangle = \pi^{-\frac{n}{4}} \exp \left[ - \dfrac{|\mathbf{x}'-\mathbf{q}'|^2}{2} + i \mathbf{p}' \cdot \left( \mathbf{x}'-\dfrac{\mathbf{q}'}{2} \right) \right] 
	\end{cases}	\, , \label{realcohs}
\end{eqnarray}
easily obtained by substituting \eqref{decomp} into \eqref{cohs}. Since $\bzeta$ and $\bzeta^*$ in the complexified case \emph{are} complex conjugates of each other, we can change integration variables from $(\bzeta,\bzeta^*)$ to $(\mathbf{q},\mathbf{p})$, the absolute value of the jacobian determinant easily seen to be $1$. Then, keeping in mind that
\begin{equation}
	\Lambda(\mathbf{q},\mathbf{p};t) = \frac{1}{2} \left[ \left( \dfrac{\partial \mathbf{p}'(\mathbf{q},\mathbf{p};t)}{\partial \mathbf{p}} + \dfrac{\partial \mathbf{q}'(\mathbf{q},\mathbf{p};t)}{\partial \mathbf{q}} \right) + i \left( \dfrac{\partial \mathbf{p}'(\mathbf{q},\mathbf{p};t)}{\partial \mathbf{q}} - \dfrac{\partial \mathbf{q}'(\mathbf{q},\mathbf{p};t)}{\partial \mathbf{p}} \right) \right] \, , \label{detC}
\end{equation}
the final IVR is written as
\begin{equation}
	K_\text{H-K}(\mathbf{x}',\mathbf{x};t) = N \int d\mathbf{q} \, d\mathbf{p} \, \sqrt{ \det \Lambda(\mathbf{q},\mathbf{p};t) } \, e^{ i S_\text{W} (t)}  \langle \mathbf{x}' \vert \bzeta'(\mathbf{q},\mathbf{p};t) \rangle \langle \bzeta^*(\mathbf{q},\mathbf{p}) | \mathbf{x} \rangle	 \, . \label{preHK}
\end{equation}
Here, the coherent states are given by \eqref{realcohs}, the phase by \eqref{genqq} and the determinant by \eqref{detC}. Normalization in this case sets $N = (2\pi)^{-n}$, and all primed variables are evolved by the hamiltonian flow as functions of $\mathbf{q}\,,\mathbf{p}$ and $t$. This IVR is known as the \emph{Herman-Kluk} (H-K) propagator \cite{Herman1984}. 

The expression in \eqref{preHK} becomes more recognizable after a simple manipulation of its generating function. We start by explicitly de-complexifying \eqref{genpop}, \emph{i.e.}
\begin{equation}
	S_{\text{W}, \mathbb{C}} (t) = \int_0^t d\tau \left[ \frac{i}{2} \left( \bzeta \cdot \dot{\bzeta}^* - \bzeta^* \cdot \dot{\bzeta} \right) - H_\mathbb{C}(\bzeta,\bzeta^*) \right] = \int_0^t d\tau \left[ \left( \frac{\mathbf{p} \cdot \dot{\mathbf{q}} - \mathbf{q} \cdot \dot{\mathbf{p}}}{2} \right) - H(\mathbf{q},\mathbf{p}) \right]  \, ;
\end{equation} 
and since $\mathbf{q} \cdot \dot{\mathbf{p}} = d(\mathbf{q} \cdot \mathbf{p})/dt - \mathbf{p} \cdot \dot{\mathbf{q}}$, we can rewrite the action above as a function of the extended position generating function as
\begin{equation}
	\quad S_\text{W}(t)	= S(\mathbf{q}',\mathbf{q};t) - \frac{1}{2} \left( \mathbf{q}' \cdot \mathbf{p}' - \mathbf{q} \cdot \mathbf{p} \right) \, .
\end{equation} 
When the expression above enters the complex exponential, the dot products sum to the phases in the Schr\"odinger coherent states and disfigure them, resulting in 
\begin{equation}
	K_\text{H-K}(\mathbf{x}',\mathbf{x};t) = (2\pi)^{-\frac{n}{2}} \int d\mathbf{q} \, d\mathbf{p} \, \sqrt{ \det \Lambda(\mathbf{q},\mathbf{p};t) } \, e^{ i S(\mathbf{q}',\mathbf{q};t) }  \langle \mathbf{x}' \vert \bbeta'(\mathbf{q},\mathbf{p};t) \rangle \langle \bbeta^*(\mathbf{q},\mathbf{p}) | \mathbf{x} \rangle \, ,
\end{equation}
with the integral kernels given by the gaussians
\begin{eqnarray}	
\begin{cases}
	\,\,\, \langle \bbeta^*(\mathbf{q},\mathbf{p}) | \mathbf{x} \rangle \,\,= \pi^{-\frac{n}{4}} \exp \left[ - \dfrac{|\mathbf{x}-\mathbf{q}|^2}{2} - i \mathbf{p} \cdot \left( \mathbf{x} - \mathbf{q}\right) \right]  \\[8pt]
	\langle \mathbf{x}' \vert \bbeta'(\mathbf{q},\mathbf{p};t) \rangle = \pi^{-\frac{n}{4}} \exp \left[ - \dfrac{|\mathbf{x}'-\mathbf{q}'|^2}{2} + i \mathbf{p}' \cdot \left( \mathbf{x}'- \mathbf{q}' \right) \right] 
	\end{cases}	\, , 
\end{eqnarray}
This is the original form of the H-K propagator as discovered by Herman and Kluk \cite{Herman1984}, and remains the most popular one. However, if one interprets the kernels above as Klauder coherent states, the profound connection this propagator has with the S-B representation is lost. Even more importantly, the fact that the action entering the H-K propagator is Weyl-ordered is also eclipsed \cite{Grossmann1998}. 

It is also common to encounter the H-K propagator expressed as a function of coherent states that depend on a real parameter, associated to its width. Everything we have done generalizes to this case by simply rescaling the complexification map (\emph{e.g.}~as in \cite{Fierro2006}) as a function of a free parameter: This will modify the H-K's pre-factor and the Schr\"odinger states accordingly. However, the matter of why a particular width works better than another appears to be specific to the particular problem at hand. 

\subsection{Caustics and the failure of semiclassical propagation}\label{subsec:fail}

The matter of whether or not the pre-factor inversion taking place in the IVR \eqref{IVR} is an improvement over the vV-G propagator was left unanswered and is now placed under scrutiny. We shall focus on the semiclassical impact of being near or far a caustic, but not exactly on it, since in this case we already know the answer: Caustics do not contribute in any way. This can be seen by noticing that in vV-G they have to be manually excluded, and in the IVR they are assigned a null pre-factor. The end result is clearly the same, showing that even though the IVR does not diverge, it is still impacted by caustics.

The caustic condition of singular $B$ matrices in \ref{subsec:lin} is identical in the case of general hamiltonian flows, but now the linearized dynamics is given by the monodromy matrix and a caustic happens when its $B$-equivalent, namely the monodromy component appearing in the vV-G propagator's pre-factor, vanishes. As discussed earlier, when this happens, \eqref{vV} and \eqref{IVR} follow
\begin{equation}
\frac{\partial^2 S(\mathbf{q}',\mathbf{q};t)}{\partial \mathbf{q}' \, \partial \mathbf{q}} \longrightarrow 0 \quad \Longrightarrow \quad K_\text{vV-G}(\mathbf{q},\mathbf{q}';t) \longrightarrow \infty \, , \quad K_\text{IVR}(\mathbf{q},\mathbf{q}';t) \longrightarrow 0 \, . \label{fail1}
\end{equation}  
Since we are in a Wentzel-Kramers-Brillouin (WKB) scenario, phases are assumed to be stationary, restricting the actions to regions in which they vary slowly. As we near a caustic the condition above goes one step further and tells us that the action changes even less, since their second derivatives also vanish \cite{Schulman1994}: The closest we are to a caustic, the slower the oscillations in the integrands of both the vV-G propagator and its IVR. As we move away from the caustic, the pre-factor in vV-G starts to decrease, which helps muffling the fast oscillations that its complex exponential begins to develop. The IVR develops the opposite behavior, assigning small contributions near caustics and huge ones as we move away from them. The consequence is that the IVR becomes both highly oscillatory and numerically large in regions where vV-G is small. Since vV-G relies on the root-search for selecting what trajectories to be included, the oscillatory behavior of its complex exponential is not a problem, as the trajectories corresponding to momenta in oscillatory regions are handpicked. The IVR, however, integrates over initial momenta and relies on the Riemann-Lebesgue lemma to annihilate the contributions emanating from unimportant trajectories. The problem is that, as we now know, these regions of fast oscillatory behavior are assigned very large numerical values by the IVR, and expecting them to cancel perfectly is far-fetched. 

The obvious way of dealing with the oscillatory numerical errors prone to appear when using the IVR is by employing very large momentum grids: They will help canceling contributions from unimportant trajectories far from caustics, while increasing the number of contributions from important trajectories in the neighborhood of caustics (which need some help due to their small pre-factors). We then see that the difficulties of the root-search in \eqref{vV} are transformed into convergence problems in \eqref{IVR}. By using the powerful computers nowadays available, however, it is generally easier to increase grid sizes than to solve the root-search. This is especially true due to the existence of many numerical methods specialized in oscillatory integrals, together with a continuing interest by the chemical community to find strategies that help improving the convergence of IVRs in practical applications. 

In the beginning of this subsection we stated that caustics do not contribute to neither \eqref{vV} nor \eqref{IVR}, which is true. The severest problem is them \emph{failing} to contribute. To see this, suppose phase space is filled with caustics, a situation that usually takes place at long propagation times in both integrable and chaotic systems \cite{Schulman1994}. This increases the probability of trajectories falling on them, such that a possible contribution from a root-momentum that would be included in the vV-G propagator ends up having to be removed. This becomes an increasingly likely event as time grows, causing more and more contributions that would enter the semiclassical sum to be left out. The missing terms, of course, would be fundamental to conserve normalization, such that we can expect both the vV-G and its IVR to lose normalization as time grows. A time-threshold must then exist in which the density of caustics becomes large enough for a complete failure of the vV-G propagator and its IVR, and in the next section we will see that this problem is rendered even more serious due to the phenomenon of \emph{caustic stickiness}. 

We now see that the H-K propagator behaves in a markedly different way from vV-G and its cousins. In fact, not a single aspect of the mechanisms for semiclassical failure described above applies to it, since their backbone is the caustic singularities it does not possess. The mechanisms for the failure of H-K when applied to integrable systems are, as far as we know, still unclear. However, it suffers from the same problem as all IVRs with regard to its pre-factor possibly diverging when dealing with chaotic dynamics. The exponential separation of trajectories that begin infinitesimally close gives birth to positive Lyapunov exponents, such that \eqref{fail1} is reversed to 
\begin{equation}
	\frac{\partial^2 S(\mathbf{q}',\mathbf{q};t)}{\partial \mathbf{q}' \, \partial \mathbf{q}} \longrightarrow \infty \Longrightarrow K_\text{vV-G}(\mathbf{q}, \mathbf{q}';t) \longrightarrow 0 \, , \,\, K_\text{IVR}(\mathbf{q},\mathbf{q}';t) \longrightarrow \infty \longleftarrow K_\text{HK}(\mathbf{q},\mathbf{q}';t) \, .\label{fail2}
\end{equation}  
We already know that the trajectories that are far from caustics generate small contributions to the vV-G propagator, but for the case of chaotic dynamics, the situation is extreme: If a trajectory is both chaotic and far from a caustic, its contribution is inversely proportional to its rate of growth (which is exponential!). Now, for the IVR and H-K, the only hope is annihilating the oscillatory terms associated to chaotic trajectories in integration. Since in this case they both have diverging pre-factors, this might be hopeless. Despite these problems, the H-K has been found to perform unexpectedly well for situations of soft chaos, in which phase space is populated by both regular and chaotic dynamics \cite{Schoendorff1998,Maitra2000,Lando2019-2}. The reason for this might be that the main contributions come from the regular trajectories, since, as states earlier, for chaotic trajectories the pre-factors diverge very fast. Indeed, an artificial erasure of chaos in a strongly chaotic system was shown to provide better semiclassical results than the ones obtained from the system's original chaotic dynamics \cite{Lando2019-3}.

\section{Numerical simulations}\label{sec:kerr}

We now begin the second half of this manuscript, which concerns numerical aspects of the vV-G propagator, its IVR and the H-K propagator. The homogeneous Kerr system, which we chose as laboratory, is integrable and does not display the intrinsic complications present in chaotic systems. Nevertheless, it does contain a quite intricate caustic web, and since we have no reason to suspect the caustics in integrable systems to be any different from the ones in chaotic systems, many aspects observed for regular dynamics should migrate unmodified to the chaotic case \cite{Schulman1994}. 

\subsection{The homogeneous Kerr system}\label{subsec:kerr}

In order to test semiclassical propagators we need a system complex enough to have caustics, but as simple as possible for numerical computations to be performed quickly. The Simple Harmonic Oscillator (SHO) is an example of such a system, but since its hamiltonian is quadratic we are stuck with the linear theory developed in \ref{sec:lin}, \emph{i.e.}~the classical and quantum evolutions are identical. A small modification of the SHO turns out to be ideal as a toy model for semiclassical techniques, since it presents an intricate web of caustics and both its classical and quantum equations of motion have analytical solutions, with quantum dynamics markedly different from its classical counterpart. The \emph{homogenous Kerr system} (or simply \emph{Kerr system}), which we have just described, is obtained from the simple 1-dimensional hamiltonian
\begin{equation}
	H_\text{Kerr}(q,p) = (p^2 + q^2)^2 \, , \label{Ckerr}
\end{equation}  
which is nothing but a rescaled SHO squared. Writing the real Hamilton equations in \eqref{compham} and dividing $\dot{p}$ by $\dot{q}$, we see that the differential equation for the flow's geometry is the same as in the SHO, \emph{i.e.}~the orbits are circles. Thus, the distance from the origin is conserved for all orbits and the flow is given by  
\begin{equation}
	\begin{cases}
	q'(q,p,t) = q \cos \left[ \omega(q,p) \, t \right] + p \sin \left[ \omega(q,p) \, t \right] \\[4pt]
	p'(q,p,t) = p \cos \left[ \omega(q,p) \, t \right] - q \sin \left[ \omega(q,p) \, t \right]
	\end{cases} \, , \quad \omega(q,p) = 4(q^2+p^2) \, ,\label{flow}
\end{equation}
which is very similar to the SHO. The only difference is that, whilst the angular velocity is the same for all orbits in the SHO, in the Kerr system it is conserved \emph{per orbit}, but monotonically increasing as a function of the distance from the origin. Using \eqref{genqq}, it is also easy to show that the classical action for the Kerr system as obtained from the flow above is given by
\begin{equation}
	S_\text{Kerr}(q'(q,p),q,t) = \frac{1}{4} \left\{ \omega(q,p) \, t + 2 \, p\, q ( \cos \left[ \omega(q,p) \, t \right] - 1) + (p^2-q^2) \sin \left[ \omega(q,p) \, t \right]  \right\} \, , \label{genkerr}
\end{equation}
while the symmetric action in \eqref{genpop} is just
\begin{equation}
	S_\text{W}(t) = (p^2+q^2)^2 t \, .
\end{equation}

Moving to the quantum realm, the canonical quantization of the Kerr hamiltonian in \eqref{Ckerr}, namely
\begin{equation}
	\hat{H}_\text{Kerr}(\hat{q},\hat{p}) = (\hat{p}^2 + \hat{q}^2)^2 \, , \label{Qkerrqp}
\end{equation}  
is straightforward and presents no ordering problems. However, just as in the SHO case, it is considerably simpler when expressed as a function of complexified variables. Here, the position and momentum operators are substituted by $\hat{p} = (\hat{\zeta}^* + \hat{\zeta})/\sqrt{2}$ and $\hat{q} = i(\hat{\zeta}^* - \hat{\zeta})/\sqrt{2}$, with the number operator given by $\hat{n} = \hat{\zeta}^* \hat{\zeta}$. In theses variables, the hamiltonian in \eqref{Qkerrqp} is brought to
\begin{equation}
	\hat{H}_\text{Kerr} (\hat{n}) = \left( 2 \hat{n} + \hat{1} \right)^2 \, , \label{QkerrC}
\end{equation}  
which is yet again our familiar squared and rescaled SHO, and as an exclusive function of $\hat{n}$ it also shares its eigenfunctions, the Fock states $|n\rangle$, with it. This is particularly useful for calculating the time-evolution of arbitrary states, since we can decompose them in the complete Fock basis $\{|n\rangle\}$ and deal with the evolution operator as a number. In particular, the eigenfunctions of the annihilation operator $\hat{\zeta}^*$ in the Fock basis are known from basic quantum mechanics \cite{BallentineBook} to be
\begin{equation}
	\vert \alpha \rangle = e^{-\frac{\vert \alpha \vert^2}{2}} \sum_{n=0}^\infty \frac{\alpha^n}{\sqrt{n!}} \vert n \rangle \, , \label{coher}
\end{equation}
representing Klauder coherent states centered at $(q,p) = \sqrt{2} (\Re(\alpha),\Im(\alpha))$. Notice these states are not the same as the Schr\"odinger states used in H-K, since they are normalized \cite{GazeauBook,KlauderBook}. The normalization is required because these will serve as initial states for quantum propagation and we want the standard probabilistic interpretation of quantum mechanics to remain valid. 

In one dimension, the position representation of \eqref{coher} is 
\begin{equation}
	\langle x | \alpha \rangle = \pi^{-\frac{1}{4}} \exp \left\{ -\frac{ \left[ x - \Re(\alpha) \right]^2 }{2} + i \Im(\alpha) \left[ x - \Re(\alpha) \right]  \right\} \, ,
\end{equation}
where we have rescaled $\alpha \mapsto \alpha/\sqrt{2}$ in order to center the state at $(q,p) = (\Re(\alpha),\Im(\alpha))$. The time-evolution of this state in the Kerr system, namely
\begin{equation}
	\vert \alpha(t) \rangle = \hat{U}_\text{Kerr}(t) \vert \alpha \rangle = e^{- i t \hat{H}_\text{Kerr}} \vert \alpha\rangle \, , \label{qflow}
\end{equation}
has been shown to be exact for times of the form 
\begin{equation}
	t = \left( \frac{2a}{b} \right) T_\text{rev} \, , \quad  T_\text{rev} = \frac{\pi}{4} \, , \quad a, b \in \mathbb{Z} \, , \quad b \quad \text{odd} \, ,  \label{rev}
\end{equation}
where $T_\text{rev}$ is known as the \emph{revival time} for this system \cite{Yurke1986,Stoler1986}. The final state is then represented as a superposition of $b$ coherent states placed symmetrically around the origin. Multiples such as $T_\text{rev}/8$ and $T_\text{rev}/16$ are especially striking due to the emergence of \emph{fractional revival patterns}, in which the evolved state is formed by the superposition of 2 (a Schr{\"o}dinger's cat \cite{Robinett2004}) or 4 (a compass state \cite{Zurek2001}) coherent states, respectively. As the name suggests, the initial state is completely recovered at $t=T_\text{rev}$ up to a global phase.

\begin{figure}
	\includegraphics[width=\linewidth]{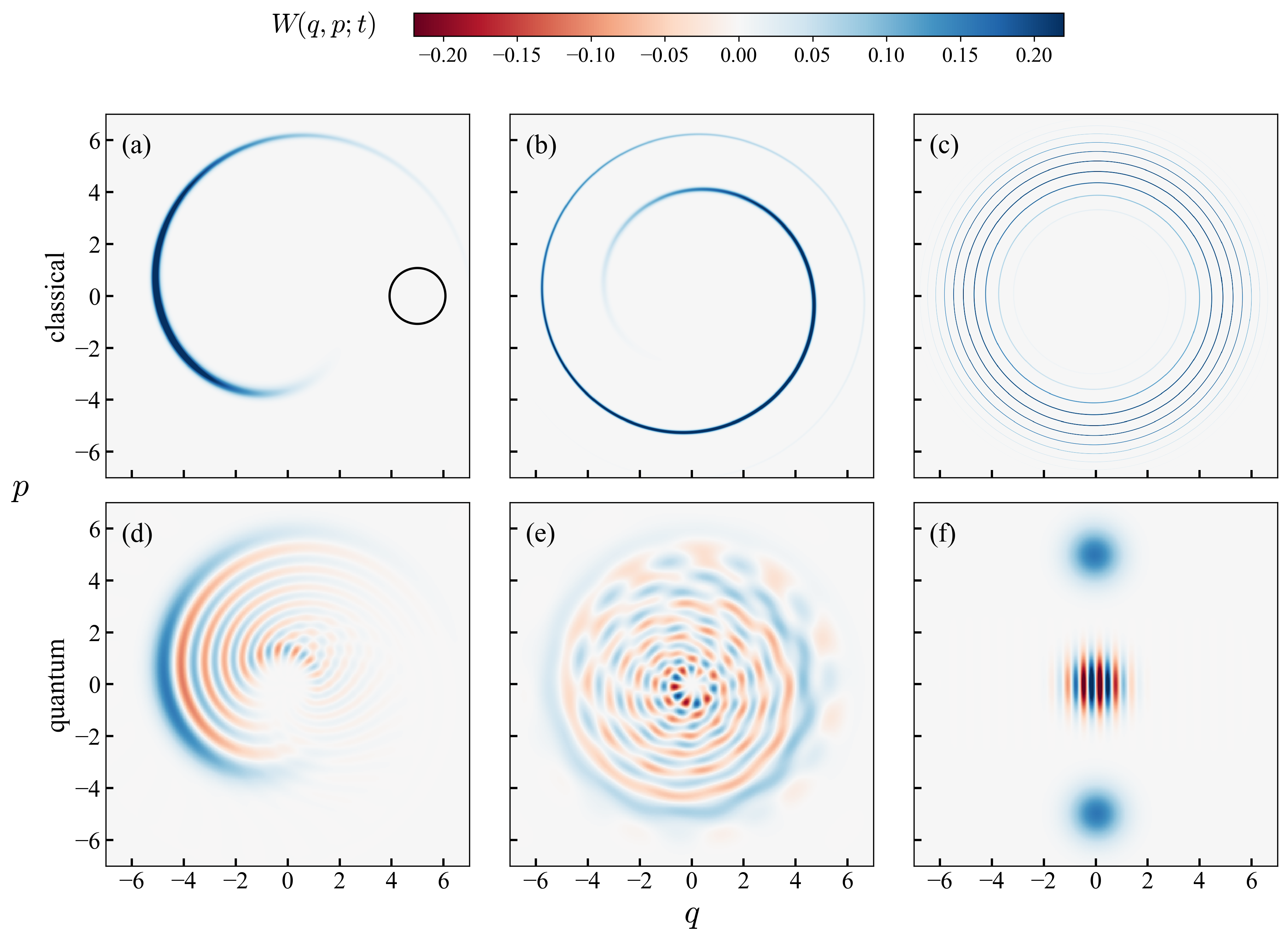}
	\caption{The TWA in \eqref{TWA} for a coherent state initially centered at $(\Re(\alpha)=5,\Im(\alpha)=0)$ and evolved by the flow \eqref{flow} (upper row), together with the exact Wigner function \eqref{wig} for the quantum evolution \eqref{qflow} (bottom row). The contour in \textbf{(a)} marks the location of the initial state. The time values for the panels are \textbf{(a, d)} $t_1 = (62/1571)T_\text{rev} \approx 0.031$; \textbf{(b, e)} $t_2 = (142/1571)T_\text{rev} \approx 0.071$ and \textbf{(c, f)} $t_3 = (786/1571)T_\text{rev} \approx \pi/8$. The TWA shows the deformation suffered by the initial state due to the outward increase of angular velocities, eventually transforming it into a barely visible filament. The situation is markedly different for the quantum evolution, which does present a classical footprint for \textbf{(d)} and \textbf{(e)}, but does not resemble its classical equivalent at all for the cat revival in \textbf{(f)}. Note that at $t_1=0.031$ the Wigner function's tail has performed a [barely visible] full revolution around the origin.}
	\label{fig:flow}
\end{figure}

The Kerr system has been already investigated in depth \cite{Yurke1986,Lando2019,Stoler1986,Toscano2009}, but its dynamics in phase space is worthwhile revisiting due to its fascinating geometry. We start by noticing that a classical phase space distribution under the action of the Kerr flow \eqref{flow} will simultaneously revolve around the origin and be deformed into a filament, since outer points move faster than inner points. In particular, the 1-dimensional \emph{Wigner transform} of a time-dependent wave function $\langle q | \psi(t) \rangle$, given by
\begin{equation}
	W (q,p;t) = \pi^{-1} \int d\tilde q \, \langle q + \tilde q | \psi(t) \rangle \langle \psi(t) | q-\tilde q \rangle  e^{-2i \, \tilde{q} \, p}\,, \label{wig}
\end{equation}
provides a classical phase space distribution from its \emph{Truncated Wigner Approximation} (TWA), which corresponds to the $\mathcal{O}(\hbar^0)$ term in an $\hbar$-expansion of Moyal's equation \cite{Groenewold1946,Moyal1949,LandoThesis,Titimbo2020}. For the initial coherent state in \eqref{realcohs} the TWA is just
\begin{equation}
W_\text{TWA}(q,p;t) = \pi^{-1} \exp \left\{ - \left[ q'(q,p,-t) - \Re(\alpha) \right]^2 - \left[ p'(q,p,-t) - \Im(\alpha) \right]^2 \right\} \, , \label{TWA}
\end{equation}
\emph{i.e.}~a phase space gaussian with classically evolving coordinates. The negative time in the r.h.s.~is not a typo, being instead a fundamental ingredient in order for the TWA to evolve as a classical phase space distribution, obeying the Liouville equation \cite{LandoThesis}. In Fig.~\ref{fig:flow} we display the exact Wigner function in \eqref{wig} and its TWA for a coherent state propagated by the Kerr dynamics for some selected times.

Fig.~\ref{fig:flow} inspires pessimism with regard to a semiclassical approximation being able to reproduce quantum evolution, especially considering that its filamentary classical backbone gets thinner and thinner as time evolves (although its area obviously remains constant, by Liouville's theorem). This expectation was proven wrong in at least three occasions: In \cite{Toscano2009} it was was shown that a careful application of the vV-G propagator was successful in reproducing the evolved wave functions for more than one revival time; in \cite{Grossmann2016} the H-K propagator was used to model a 0-dimensional Bose-Hubbard chain, for which the hamiltonian is given by the slightly different (yet dynamically identical) expression; and in \cite{Lando2019} a value representation using final instead of initial values, proposed first in \cite{Ozorio2013}, was able to reproduce quantum dynamics with calculations performed directly in phase space. Since we know that these semiclassical analyses of the Kerr system were successful, we can be sure that semiclassical methods are supposed to work for this system.

It is fundamental to keep in mind that the previous analysis \cite{Toscano2009} of the Kerr system using the vV-G propagator used a series of approximations to obtain accurate results, such as filtering trajectories and approximating the action up to second order. Here, however, we are not interested in obtaining accurate results, but in providing fair comparisons between all semiclassical propagators, which are calculated in the same grids and with the same number of trajectories unless explicitly stated. Our objective is to use the methods in the most plug-and-play possible manner, without any approximations, trajectory focusing or optimization. This is only possible because all classical objects used by the semiclassical propagators for the Kerr system are obtained from analytical calculations, \emph{except} for the root-trajectories, the Maslov indexes and the branch changes. In \ref{App:C} we show how the error in the root-search of vV-G can be made equivalent to the numerical one in the flow, and since we use the same algorithm to calculate Maslov indexes and branch changes, the error in both objects is the same. It is also worthwhile to mention that, since branch changes and Maslov indexes are both obtained from a comparison algorithm, there is no numerical advantage at all in the absence of an explicit index in H-K: Guaranteeing the continuity of its pre-factor is a numerically identical process to counting caustics in vV-G \cite{Kay1994,SwensonThesis}.

\subsection{A quick look at pre-factors}\label{subsec:pre}

\begin{figure}
	\includegraphics[width=\linewidth]{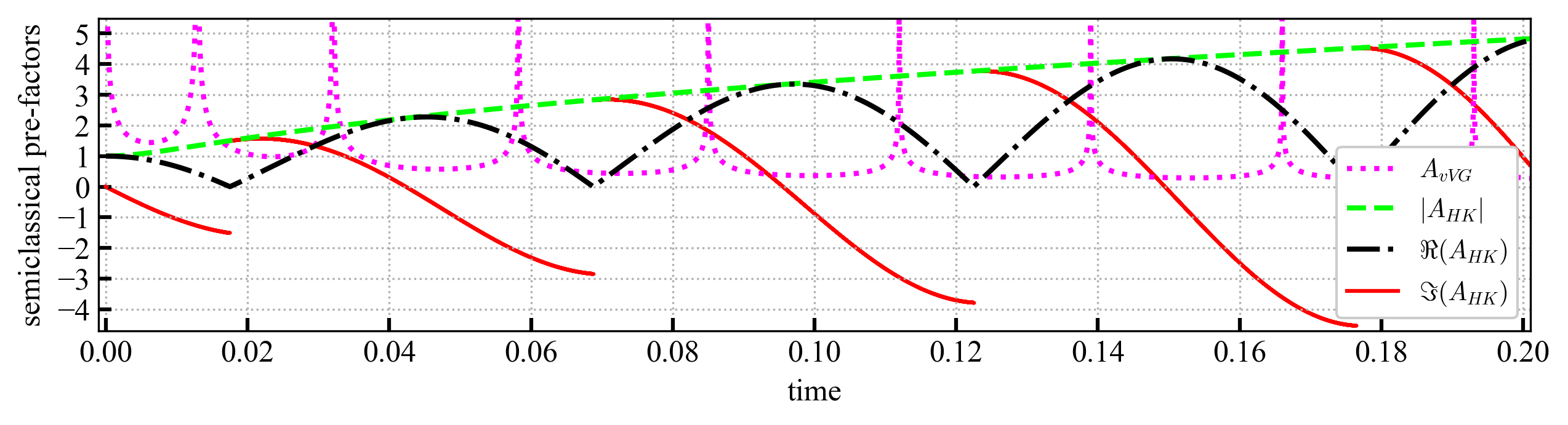}
	\caption{Pre-factors in the vV-G and H-K propagators, given respectively by \eqref{prevvg} and \eqref{prehk}, for the initial point $(q=5,p=2)$.}
	\label{fig:pres}
\end{figure}

The first semiclassical aspect we would like to investigate is the difference between the pre-factors in the vV-G and H-K propagators in \eqref{vV} and \eqref{preHK}, which in the 1-dimensional case are just
\begin{eqnarray}
	A_{vVG}(q_0,p_0;t) &= \left\vert \frac{\partial q'(q,p,t)}{\partial p} \right\vert^{-\frac{1}{2}}_{(q_0,p_0)} \label{prevvg} \\ 
	A_{HK}(q_0,p_0;t) &= \left\{ \frac{1}{2} \left[ \left( \dfrac{\partial p'(q,p;t)}{\partial p} + \dfrac{\partial q'(q,p;t)}{\partial q} \right) \right. \right. \notag \\
	&\qquad \qquad \left. \left. + i \left( \dfrac{\partial p'(q,p;t)}{\partial q} - \dfrac{\partial q'(q,p;t)}{\partial p} \right) \right] \right\}^{\frac{1}{2}}_{(q_0,p_0)} \, , \label{prehk}
\end{eqnarray}
and can be analytically obtained from the derivatives of the flow in \eqref{flow}. By choosing an arbitrary point $(q_0,p_0)$ and plotting these pre-factors as a function of time, fundamental differences between both methods can already be seen, as we display in Fig.~\ref{fig:pres}. For instance, by keeping in mind that the classical action \eqref{genkerr} inherits the flow's periodicity, we see that the asymptotic behavior of $A_{vVG}$ towards zero is a further indicative that the vV-G propagator might lose normalization as time grows; The absolute value of $A_{HK}$, on the other hand, suggests that in this case the problem is quite the opposite, with the H-K propagator risking growing too much. As is clear in the figure, however, $A_{vVG}$ goes to zero exponentially, while $A_{HK}$ grows only logarithmically, suggesting that the vV-G propagator loses normalization faster than the H-K propagator diverges. Notice that since $\partial_p q = 0$, $A_{vVG}$ has a caustic at the origin independently of the choice of initial phase-space point, as discussed in Sec.~\ref{subsec:sqb}. 

Another important aspect seen in Fig.~\ref{fig:pres} is that the caustics in vV-G do not occur at the same places as the branch changes in H-K, but the number of caustics between each branch change is always equal to two, as demonstrated by Kay in one of the earlier investigations on IVRs \cite{Kay1994}. It is also clear that the absolute value of the H-K pre-factor is always larger than 1, as proved in \ref{App:A}, and that the branch changes happen when the real part of the pre-factor hits zero and its imaginary part changes sign. We do not show the IVR pre-factor in \eqref{IVR} in order to avoid convoluting Fig.~\ref{fig:pres}, but since it is the inverse of vV-G's it is quite clear that it will grow in time and tend to zero as caustics are approached.

\subsection{Implementing semiclassical propagation}\label{subsec:impl}

We now move to the implementation of the semiclassical propagators in \eqref{vV} and \eqref{preHK}. Since the Kerr system is integrable, the root-search required by vV-G is not particularly challenging, as trajectory multiplicity can be easily dealt with as described in \ref{App:C}. For vV-G, we do not avoid caustics in any way, running over their neighborhoods and divergent points \footnote{Their neighborhoods, of course, enter the calculations in full. It is only the proper caustic, the $\infty$, that is automatically removed by the compiler in the calculations. All coding is done in the Julia programming language.}. Naturally, the process of obtaining the roots can be optimized by grid-focusing and other strategies that depend on the form of the initial state, \emph{e.g.}~if one is interested in propagating a coherent state, all trajectories with initial positions lying outside the initial coherent state can be dismissed in the calculations \cite{Heller1991}. Here, however, we do not wish to be limited to propagating wave functions, since the semiclassical propagators themselves offer a striking visual comparison. This requires us to root-search everywhere for initial momenta, such that all trajectories in our position-momentum grid are included in vV-G; For the case of H-K the whole grid is used for integration. 

\begin{figure}
	\includegraphics[width=\linewidth]{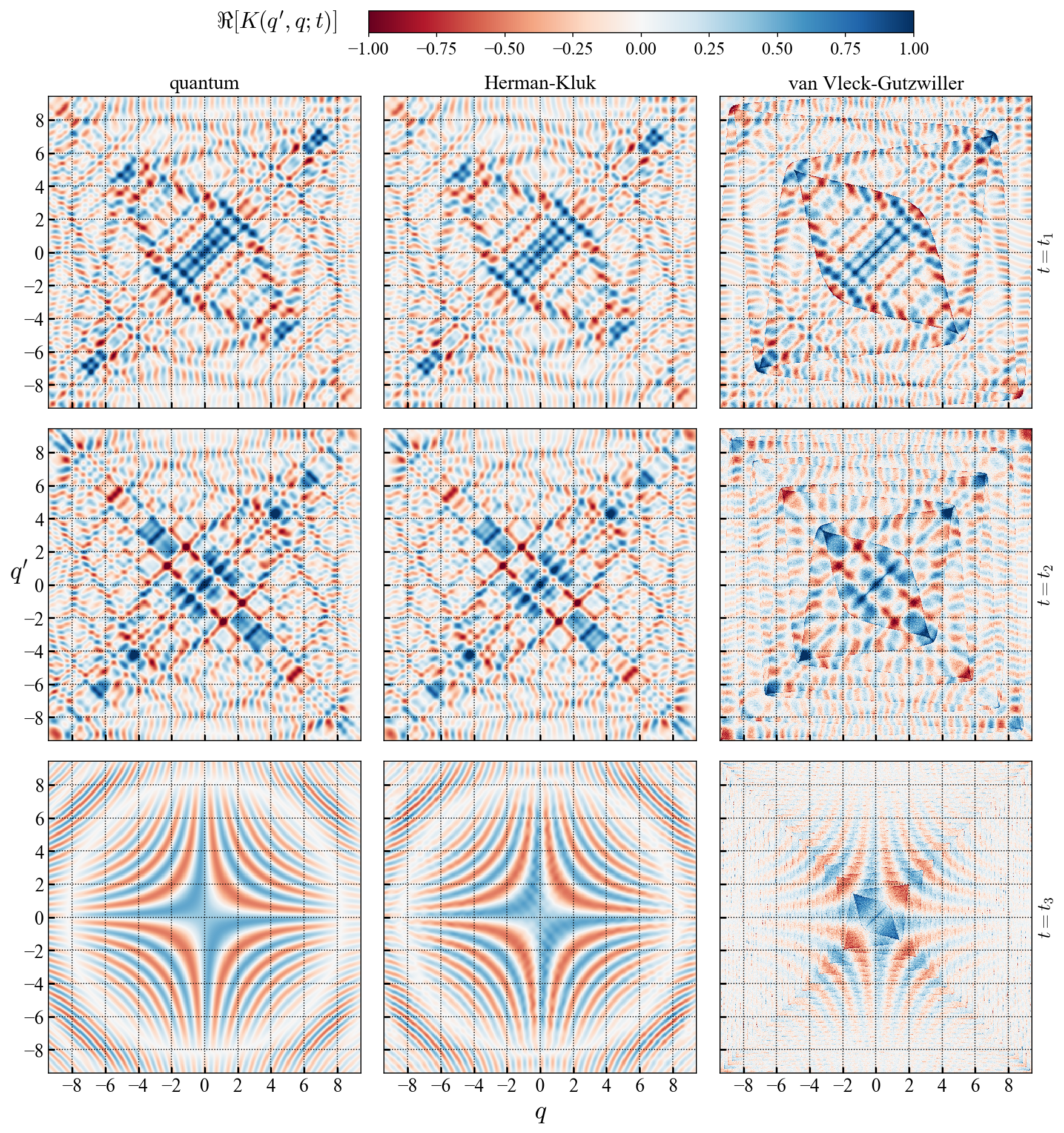}
	\caption{Real parts of the quantum \eqref{Kqu}, vV-G \eqref{vV} and H-K \eqref{preHK} propagators for the Kerr system. The imaginary parts have identical accuracy. The time values are the same as in Fig.~\ref{fig:flow}. All propagators are calculated on the same $501 \times 501$ position-momentum grid, ranging from $-3\pi$ to $3\pi$. All root-trajectories in the grid are included in vV-G except for the ones with infinite pre-factor, which are automatically excluded by the compiler.}
	\label{fig:Ks}
\end{figure}

\begin{figure}
	\includegraphics[width=\linewidth]{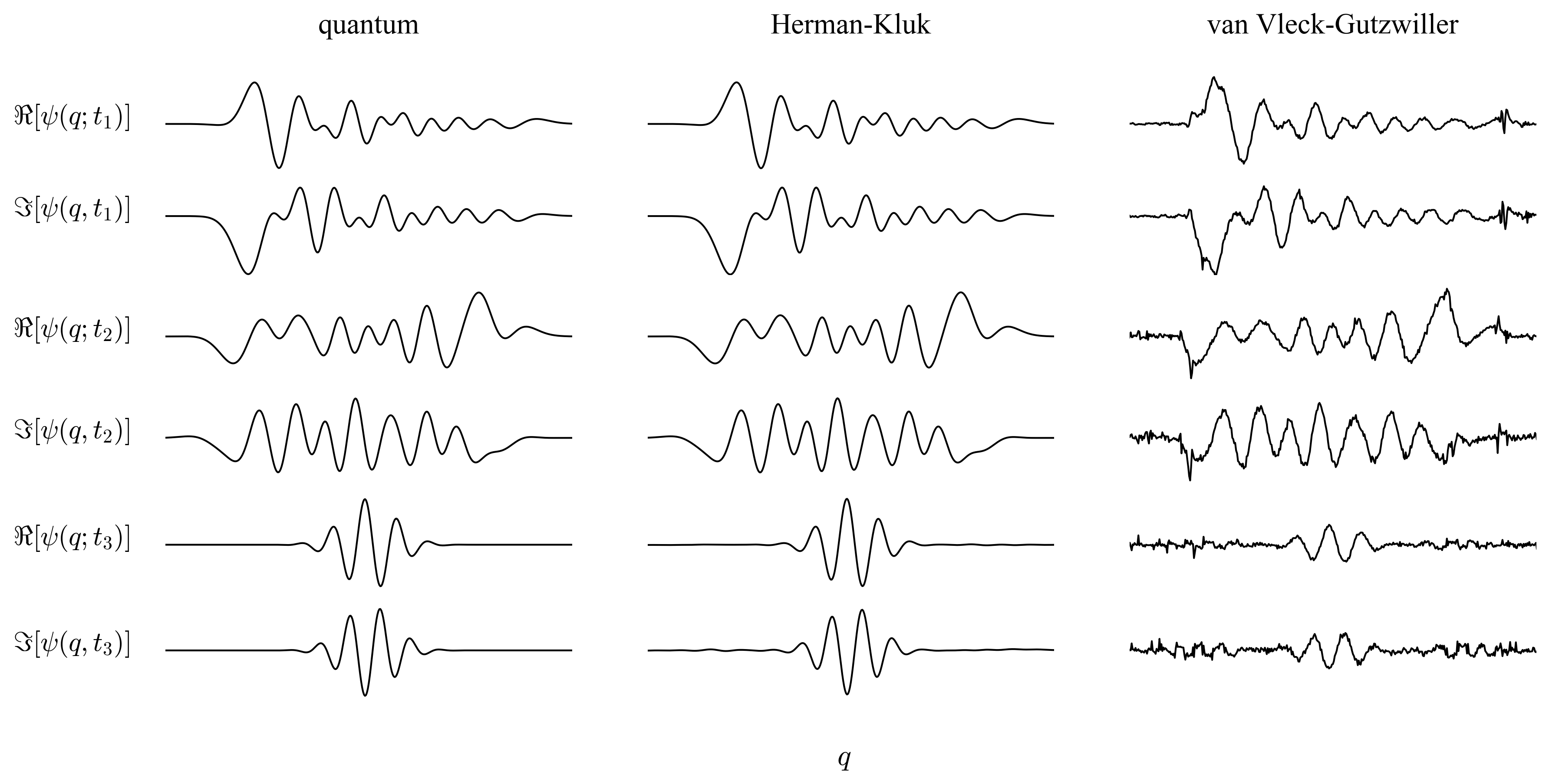}
	\caption{Wave functions for an initial coherent state centered at $(\Re(\alpha)=5,\Im(\alpha)=0)$, the same as in Fig.~\ref{fig:flow}, integrated against the propagators in Fig.~\ref{fig:Ks}.}
	\label{fig:waves}
\end{figure}

Evidently, comparing semiclassical results without their quantum equivalent is certainly lacking in comprehensiveness. To calculate the quantum propagator for the Kerr hamiltonian, we insert a Klauder coherent state projector in the expression for the position propagator:
\begin{equation}
K_\text{Kerr}(q',q;t) = \pi^{-n} \int d\zeta \, \langle q' | \hat{U}_\text{Kerr}(t) | \alpha \rangle \langle \alpha |q \rangle \, .
\end{equation}
Since the evolution of coherent states is exact for the times in \eqref{rev}, by writing the complex measure explicitly we have the exact quantum propagator
\begin{equation}
K_\text{Kerr}(q',q;t) = (2\pi)^{-1} \int d\Re(\alpha)\,d\Im(\alpha) \, \langle q' | \alpha (t) \rangle \langle \alpha |q \rangle \, , \label{Kqu}
\end{equation}
with $|\alpha(t)\rangle$ as in \eqref{qflow} for the times in \eqref{rev}. A comparison of quantum, vV-G and H-K propagators can be seen in Fig.~\ref{fig:Ks} for the same time values as in Fig.~\ref{fig:flow}.

Two aspects of Fig.~\ref{fig:Ks} immediately catch the eye: The first is the astonishing accuracy of the H-K propagator, which is almost indiscernible from its exact quantum counterpart; The second is the trapezoidal structure formed by caustic submanifolds lifted to the $(q',q)$-space, visible in the vV-G propagator (for caustics in the $(q,p)$-space, see Fig.~\ref{fig:caustics}). As discussed earlier, quantum propagation is smooth, so that the web of caustics appearing in vV-G has no equivalent in the quantum world and is absent in the caustic-free H-K. Indeed, for $t=t_3$ this web is so dense that the final propagator goes to zero in the outskirts of the grid, where caustics proliferate faster due larger angular frequencies and, in consequence, a higher number of zeros in the pre-factor. It is clear, however, that the vV-G propagator provides reasonable values for regions near the origin, in which the caustic web is sparser. We will soon see that caustics not only proliferate in time, but that the time spent by a trajectory when crossing a caustic is also increased. We suspect this to be an important mechanism for the failure of vV-G for long propagation times. 

Despite the visual richness of Fig.~\ref{fig:Ks}, it is important to have a more quantitative comparison between semiclassical propagation methods. The most immediate one is to use the propagators in Fig.~\ref{fig:Ks} to evolve an initial state and compare wave functions. For this we choose the same initial coherent state of Fig.~\ref{fig:flow}, and the result is presented in Fig.~\ref{fig:waves}. Again, the H-K wave functions are indiscernible from the exact quantum ones, while the vV-G wave functions reflect the instability of their respective propagators. Indeed, the cat-state wave function using vV-G is fading away due to normalization loss, a phenomenon also observed in the Wigner functions in \cite{Lando2019}, which required renormalization -- a quite common procedure in the field of quantum chaos. The H-K propagator, however, has already been credited with conserving its normalization for very long times \cite{Herman1986,Miller2011}, a point Fig.~\ref{fig:waves} confirms. Normalization shall be explored more deeply in Sec.~\ref{subsec:caus}.

\subsection{Autocorrelation functions}\label{subsec:auto}

\begin{figure}
	\includegraphics[width=\linewidth]{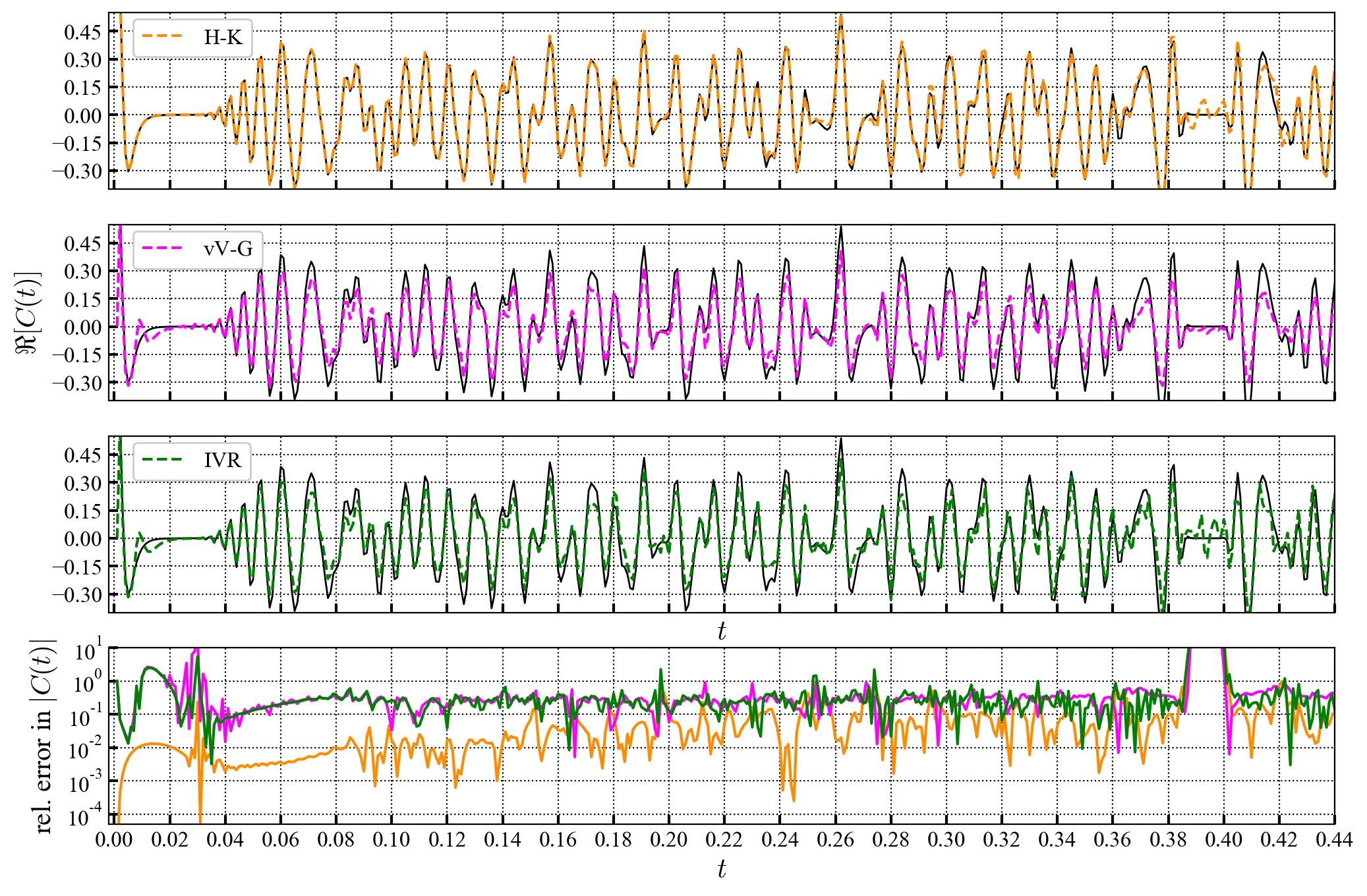}
	\caption{Real part of the autocorrelation function for the same initial coherent state used in Figs.~\ref{fig:flow} and \ref{fig:waves}, together with the relative error in the absolute value, given by $|1 - |C_\text{SC}/C_\text{QU}||$, where $C_\text{SC}$ is a semiclassical result and $C_\text{QU}$ the quantum one. The black curve in the real parts is the exact quantum autocorrelation, and the colored lines are the results obtained from each semiclassical method in the plot's legend. To calculate the propagation using vV-G and H-K we used a $201 \times 201$ position-momentum grid, from $-3\pi$ to $3\pi$, corresponding to around 3.200 points falling on the initial coherent state. The IVR does not converge on this grid, and we increase it to $1001 \times 1001$, also from $-3\pi$ to $3\pi$, for which around 80.000 points fall on the initial state. The IVR does converge with sparser grids, but we choose a dense one to make sure the results are at least comparable to vV-G. The cat state fractional revival happens at $t \approx 0.393$, about which the absolute value of the autocorrelation function is symmetric (but real and imaginary parts are not, due to phases).}
	\label{fig:auto}
\end{figure}

Several important quantities in physics and chemistry are given by numbers, such that the semiclassical propagator is integrated twice and the oscillatory behavior seen \emph{e.g.}~in the vV-G wave functions of Fig.~\ref{fig:waves}, which was already an attenuation of the one in Fig.~\ref{fig:Ks}, should be muffled even further. The autocorrelation function in \eqref{auto} is an example, and in Fig.~\ref{fig:auto} we display it as obtained from H-K, vV-G, and the IVR expression in \eqref{autoIVR}. 

It is evident from Fig.~\ref{fig:auto} that all semiclassical methods used are successful in reproducing the autocorrelation function: The quantum oscillations are accurately captured and the discrete Fourier transform of the data reproduces the quantum energy spectrum perfectly, such that we choose not to show it here as this manuscript is already quite long. Fig.~\ref{fig:auto} allows us to observe several theoretical points raised in the main text, namely:

\begin{itemize}
	\item The general caustic of position representation at $t=0$ is soon followed by a tiny time regime in which vV-G and its IVR are very accurate, which is then followed by a not so accurate intermediate short-time regime which ends in a severe and general inaccuracy for all methods used, centered around $t=0.031$. 
	\item The oscillations lose amplitude for the vV-G propagator, showing that it does lose normalization as time grows. The IVR is also affected, but as we used a much denser grid to calculate it (see caption of Fig.~\ref{fig:auto}), it loses normalization more slowly.
	\item The relative error in the autocorrelation function's absolute value calculated using H-K, just as the real and imaginary parts (not shown), shows that H-K is significantly more accurate than vV-G and its IVR.
	\item The IVR result is at best equivalent to the one obtained using vV-G, despite its 25 times larger grid.  
\end{itemize}

Besides these points, some aspects of Fig.~\ref{fig:auto} were not expected. The first is that despite all propagators having huge errors centered around the cat state revival at $t\approx0.393$, the vV-G is the one that better approximates the autocorrelation, even though its corresponding wave function in Fig.~\ref{fig:waves} has already lost a great deal of normalization and is filled with oscillatory errors. This is a stark demonstration of the filtering of oscillatory behavior taking place when integrating the semiclassical propagators on their whole domains. Of course, the fact that accurate autocorrelation functions can be obtained from semiclassical approximations that provide poor wave functions does not barren their application to, obviously, calculate autocorrelation functions. If there is interest in the connections between quantum and classical physics, however, an accurate autocorrelation function is not enough: One needs more general results, such as the propagators in Fig.~\ref{fig:Ks}, which suggest that a semiclassical quantization recipe based on a caustic-free representation ties the quantum and classical worlds much more closely than one that includes the caustics. 

The second surprising aspect of Fig.~\ref{fig:auto} is the general inaccuracy of semiclassical propagation at $t=0.031$, which is related to the \emph{Ehrenfest time} $T_\text{Ehr}$ for this particular initial state (see the first column of Fig.~\ref{fig:flow}). The Ehrenfest time in this particular case can be taken as the instant at which the initial packet's centroid has performed a full revolution around the origin \cite{Raul2012,Lando2019,Toscano2009}, obtained from the requirement
\begin{equation}
q'(q,p;0) \equiv q'(q,p, T_\text{Ehr}) \quad \Longrightarrow \quad 4(q^2+p^2) T_\text{Ehr} = 2\pi \quad \Longrightarrow \quad T_\text{Ehr} = \frac{\pi}{2(q^2+p^2)} \, ,
\end{equation} 
where $q'(q,p;t)$ is given in \eqref{flow}. For the centroid of the initial packet in Fig.~\ref{fig:auto}, we have $ T_\text{Ehr} = \pi/50 \approx 0.063$. The geometrical meaning of $T_\text{Ehr}/2$ is that at this moment the Wigner function's centroid has achieved the largest distance with respect to where it began, equal to the diameter of its orbit. As we can see from Fig.~\ref{fig:flow}, this is also the time at which the Wigner function's tail has performed a full revolution. Thus, at $T_\text{Ehr}/2$, the Wigner function has for the first time covered the maximum area available for this particular initial state, defining its \emph{characteristic action} \cite{Raul2012}.  

An equivalent interpretation of $T_\text{Ehr}$ is as the instant at which the autocorrelations obtained using the classical evolution of the Wigner function, \emph{i.e.}~the TWA, and its quantum equivalent cease to agree \cite{Lando2019,LandoThesis,Heller1991}. This time, which was previously thought to be a barrier for semiclassical propagation to work properly, has been broken on a daily basis by even the simplest of methods \cite{Voros1996}. However, it is often reported that no distinguishing feature can be observed in semiclassical propagation at the Ehrenfest time, which is something we also see in Fig.~\ref{fig:auto}: There is no feature indicating anything special about $t=0.063$. For $T_\text{Ehr}/2$, however, all propagation methods are inaccurate, implying the existence of something deeper than numerical errors, caustics, or grid sizes. This feature is not observable by naked eye in the autocorrelation functions themselves and requires us to look at the relative errors, perhaps explaining why, to our knowledge, this has not been observed before. 

We note that it is also possible to define the Ehrenfest time as half the centroid's revolution, although in this case one loses connection with the separation of quantum and classical autocorrelation functions \cite{Raul2012}.

\subsection{Caustic stickiness}\label{subsec:caus}

Caustics are unavoidable in the vV-G propagator and, as mentioned in Subsec.~\ref{subsec:impl}, infinite pre-factors lead to the contributions from their respective trajectories being lost. We here describe a mechanism that renders the vV-G propagator very unlikely to work for long propagation times, due to trajectories accumulating on caustics.

\begin{figure}
	\includegraphics[width=\linewidth]{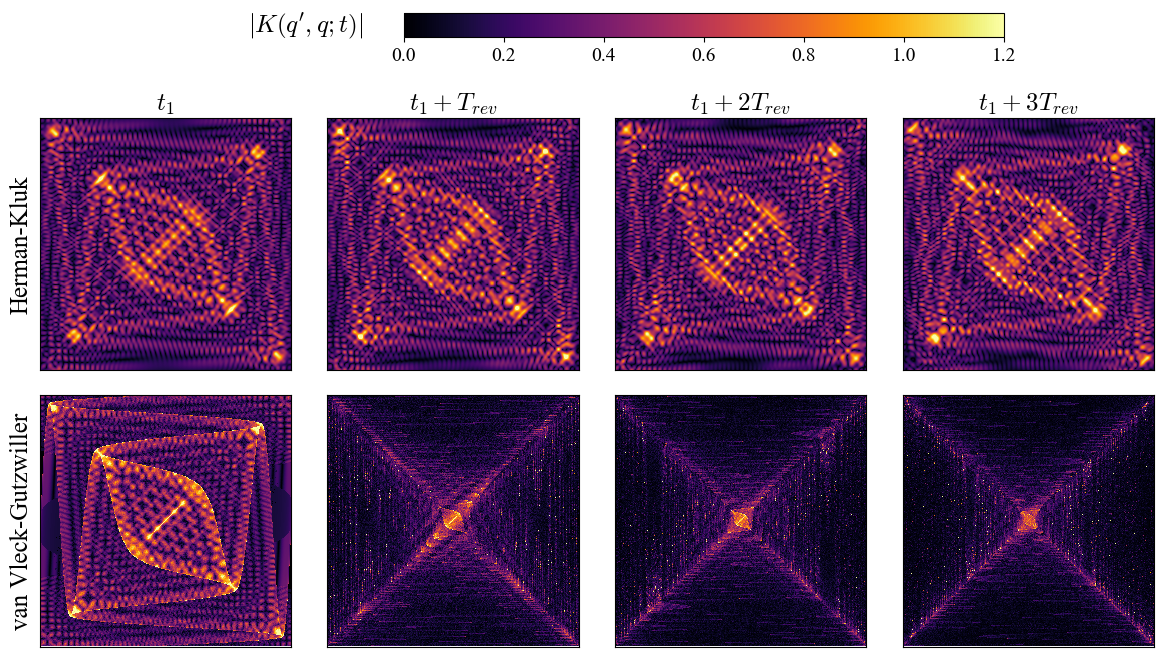}
	\caption{The absolute value of the H-K and vV-G propagators for $t_1=0.031$, where for each row one revival time is summed to $t_1$.}
	\label{fig:longtimes}
\end{figure}

\begin{figure}
	\includegraphics[width=\linewidth]{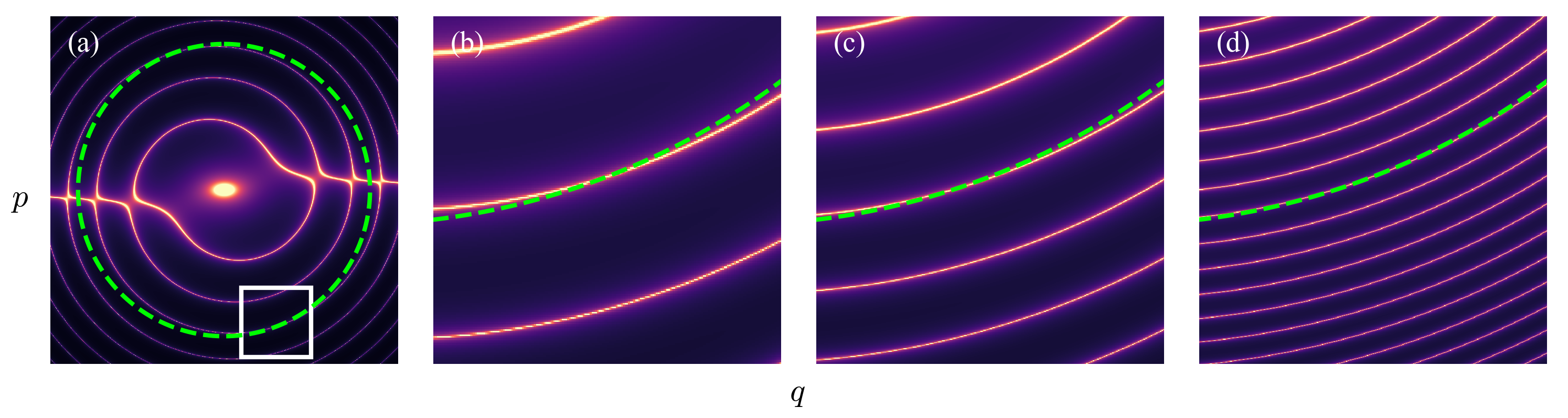}
	\caption{The phenomenon of caustic stickiness, in which trajectories keep falling on caustics as long as they are bound to the same orbit, is exemplified here. Panel \textbf{(a)} is a picture of the general caustic structure in the Kerr hamiltonian, together with a typical trajectory in bright green. The time in this panel is $t=0.117$. In \textbf{(b)} we zoom in the white square of \textbf{(a)}, where it is seen that the caustic crossing happens clearly at a point. In panel \textbf{(c)} we have $t=0.205$ and it is seen that the caustic has become more circular, such that the crossing takes longer. \textbf{(d)} Here $t=0.606$, and the whole portion of the trajectory in the white square of panel \textbf{(a)} falls on a caustic.} 
	\label{fig:caustics}
\end{figure}

In Fig.~\ref{fig:longtimes} we present the absolute value of the vV-G and H-K propagators for $t=t_1$, as in Fig.~\ref{fig:Ks}, but for each column we sum a revival time to it. The absence of caustics causes H-K to be almost unchanged, but their increasing density destroys vV-G. At first it might look perfectly possible to use immense grids and recover reasonable results, since we compensate the normalization loss by including more trajectories that do not finish on caustics. This logic can be expressed as: If my trajectory starts at $(q,p)$ and lands on a caustic at $(q',p')$, then a small perturbation in either initial positions or momenta will evade the caustic and provide a usable contribution. What we have discovered is that this logic is incorrect, as the ``spiraling'' caustic web for the Kerr system becomes more circular with the passing of time, such that all \emph{all} initial positions and momenta lying on the same orbit for that fixed time start to fall on caustics. We refer to this as \emph{caustic stickiness}.

The mechanism described above is depicted in Fig.~\ref{fig:caustics}. In panel \textbf{(a)} we show how the general structure of caustics looks like for the Kerr system, together with a typical trajectory. We then zoom on the white square to produce panel \textbf{(b)}, where we can see the trajectory crossing the caustic from up close. In panel \textbf{(c)} we go to a longer time and witness the caustic stickiness: Not only does the number of caustics increase, but they also start becoming more circular, increasing the time a trajectory remains on one of them during crossing. In panel \textbf{(d)} we are near the revival time $T_\text{rev}$, and the trajectory remains entirely on the caustic for this particular region. This has an extremely destructive effect on the final propagator, since it shows that not only a single trajectory is lost in the semiclassical sum, but instead all the trajectories ending on the green line in panel \textbf{(d)} are removed. This renders finding trajectories that do not end on caustics very hard for long times, and should also explain the unexpected null-valued regions in the Wigner functions of \cite{Lando2019}.

The times in Fig.~\ref{fig:caustics} were, of course, selected on purpose for the caustics submanifolds to support the same trajectory, but it must be kept in mind that this happens in the neighborhoods of all caustics. Besides, since caustics are also removed from the IVR by having null pre-factors, the problem migrates unaltered. It has also been pointed out that the nature of caustic crossings is	 the same for both integrable and chaotic systems \cite{Schulman1994}, such that we suspect caustic stickiness might not be restricted to the Kerr system. Although there are several interesting considerations regarding the relationship between the distribution of root-trajectories and caustics as time grows (even more stickiness!), these findings are not fundamental to this present work and shall be published elsewhere. 

\section{Discussion}\label{sec:disc}

The preference of the chemical community for the H-K propagator is completely justified, given the quality of its results, but the uneasiness of many with regard to its theoretical background has made it harder to see this propagator as what it is: An IVR for the position propagator expressed in the S-B representation. It doesn't help that this propagator did not fall prey to a consistent derivation until rather recently. Its discovery by Herman and Kluk in \cite{Herman1984} cannot be considered rigorous and received some criticism after more careful examinations failed to establish consistent links with the coherent state representation (\emph{e.g.}~\cite{Baranger2001}). Other authors, however, developed significant arguments in favor of the H-K propagator \cite{Grossmann1998,Miller2001} (and more recently \cite{Swart2009}). Some time later, Kay made use of the over-completeness of the coherent state basis to derive the H-K propagator through a series of very exhausting calculations \cite{Kay2005}. Follow up papers \cite{Fierro2006} and \cite{Ezra2006} were then the first instances where the H-K propagator was connected to complex variables by either complex WKB theory or SPAs.

Nevertheless, the fact that the H-K propagator relies on real trajectories, merely parametrized by complex variables, was never connected to its absence of caustics -- and, therefore, to its roots in the S-B space. The S-B representation of the metaplectic group, for instance, is a unitary representation of the symplectic group and is exact only because it relies exclusively on complexified, instead of complex, variables \cite{Littlejohn1986}. What we have demonstrated here is that the generalization of this to the semiclassical scenario is exactly the theoretical pillar that renders the H-K propagator caustic-free. Besides, by identifying the map taking the S-B propagator to the position one as a sequence of inverse S-B transforms, one finally understands why the coherent states in the integral kernel of the H-K propagator must follow Schr\"odinger's phase convention instead of the more obvious, normalized Klauder one: It must include the gaussian measure with which the S-B representation is equipped. A consequence is that the generating function entering the S-B propagator is unmistakably identified as the symmetric action given by the Weyl ordering rule. 

The striking distinction between representations with and without caustics is clear in the analysis of the homogeneous Kerr system. Despite its regular dynamics, the caustic submanifolds for this system are as intricate as they can be, possibly even when compared to chaotic systems. The astonishing accuracy achieved by the H-K propagator becomes even more significant if one considers that the trapezoidal caustic web, visible in the vV-G propagator in Fig.~\ref{fig:Ks}, will not go away regardless of grid size, and the accumulation of caustics as time grows will inevitably lead to general failure. As discussed, the defects in the vV-G propagator are increasingly muffled as one integrates it: The wave functions in Fig.~\ref{fig:waves} are an improvement over the propagators in Fig.~\ref{fig:Ks}, and the autocorrelation in Fig.~\ref{fig:auto} is strikingly more accurate than one would suspect by looking at the earlier figures. However, when we use propagators as integral kernels and integrate them against states, we lose touch with the fact that the links between the quantum and the classical are much clearer by looking at the propagators themselves. In this aspect, the H-K propagators displayed in Fig.~\ref{fig:Ks} are, to this day, the strongest evidence that quantum-classical connections are deeper when employing representations that are invariant with respect to the symplect group. Although here our invariant representation is the S-B one, another example is the Wigner representation of quantum mechanics, which provides arguably the most important object when looking for quantum-classical connections: The Wigner function \cite{Ozorio1998,Wigner1932}. Unfortunately, the real nature of the Wigner representation superposes its invariance with respect to the symplectic group, such that it happens to not be caustic-free. The complexification of double phases spaces \cite{Ozorio2010}, however, will possibly reward us with a caustics-free, invariant way to describe Wigner evolution. 

Another important point addressed here is that the bypassing provided by transforming a semiclassical propagator to an IVR doesn't get rid of the problems raised by caustics, since they have nothing to do with the propagator, but with the representation it uses. Thus, the loss of contributions taking place in raw propagators migrates unaltered to their IVRs, which in turn have much worse convergence than the raw propagators themselves. Although the uncomfortable process of root-searching is avoided, the inversion of pre-factors that takes place when moving to an IVR drastically increases the amplitude of oscillations already present in the raw propagator, rendering the IVR dependent on much larger integration grids than the raw propagators themselves in order to converge. Naturally, since there are no caustics in the S-B representation, this maximization of amplitudes does not happen in the H-K propagator and, to add yet another desirable numerical aspect of this method, the static and evolving coherent states in its kernel limit the integration domain to trajectories close to the one which connects their centroids. This prevents the H-K propagator to include trajectories that are far from its main stationary one and would give off mostly oscillatory errors. All these characteristics sum up to provide a semiclassical propagator that converges with very few trajectories and has minimal numerical errors.

We here also identify the impact of a characteristically classical time, namely half Ehrenfest's, in semiclassical propagation. A major mechanism for the failure of semiclassical propagators based on representations that have caustics, which we dubbed as ``caustic stickiness'', is also presented. As the density of caustics increases in phase space, whole families of trajectories (the ones lying on the same classical orbit) are lost in both vV-G and its IVR. We suspect that the loss of trajectories due to caustic stickiness explains the presence of blank arcs inside the coherent states at the fractional revivals in \cite{Lando2019}, since these have the same geometry of the caustic submanifolds in the Kerr system. Although the marginals obtained from the Wigner function can be significantly improved by using a larger grid, the arcs will never disappear from the Wigner function itself. Likewise, very large integration grids improve the results obtained \emph{from} the propagator and are useful in calculations, but the propagator itself will always reflect the caustics -- they are the footprints of working with a compromised representation. 

We do not claim to have implemented the vV-G propagator in the smartest possible manner (as in \cite{Toscano2009}), but in general such manner might not even exist. In fact, we have not implemented H-K any less crudely than we have vV-G, since we used the same integration grids and algorithms for all methods. The numerical advantage given to the position space IVR is necessary in order to achieve convergence, but IVRs of this type are not so interesting as in phase space, where they can be used to calculate the Wigner functions themselves \cite{Ozorio2013}. It could be argued that the vV-G propagator could provide better results if a larger number of trajectories were included in the sum, but we did try to increase the root-search domain and the impact was very small. Besides, we can also reverse the argument and state that the H-K propagator achieved strikingly accurate results with the same trajectories available to vV-G. In the end, we are also unable to see how including more trajectories would get rid of the problem of caustic stickiness.

Nothing in this manuscript is indicative that using IVRs, whether the position one or even the H-K propagator itself, will be fruitful for the study of hard chaos or even strong soft chaos, in which chaotic trajectories cover a larger portion of phase space than the regular ones. A comprehensive analysis of the employment of the H-K propagator to chaotic systems is still lacking, but it might indicate that vV-G is not excluded as the method of choice in this case. 

\section{Conclusion}\label{sec:conc}

We have shed new light on the Herman-Kluk propagator, which has for many years evaded proper theoretical contextualization. Its root in the Segal-Bargmann representation was shown to be the reason for its lack of caustics, which are a general feature of semiclassical propagators based on other representations. After a deep numerical investigation, this propagator's stringent success might imply that the connections between the classical and quantum realms are better established by using coherent states, an argument as old as quantum mechanics itself \cite{Schrodinger1926}. We did not, however, investigate this propagator's behavior for systems presenting chaotic dynamics, which might well be its Achilles' heel at least for the hard chaotic scenario. Nevertheless, since we have no reason to assume caustics in integrable and chaotic systems to be any different, we expect our conclusions to be generalizable to higher dimensional and/or chaotic dynamics.

\section*{Acknowledgements}

I thank Frank Gro{\ss}mann, Ranieri V.~Nery, and Steven Tomsovic for fruitful discussions. Especial gratitude is reserved to Alfredo M.~Ozorio de Almeida, for both invaluable scientific guidance and a most careful read of this manuscript.

\appendix

\section{Proof of the non-singularity of $\Lambda$}\label{App:A}

For any symplectic matrix $\mathcal{M}$ we have 
\begin{equation}
\mathcal{M}^T \mathcal{J} \mathcal{M} = \mathcal{J} \quad \Longrightarrow \quad \mathcal{M}^{-1} = \mathcal{J}^{-1} \mathcal{M}^T \mathcal{J} \, . \label{app1}
\end{equation}
Now, $\mathcal{M}$ can be obtained from the de-complexification of $\mathcal{M}_\mathbb{C}$ as $\mathcal{M} = \mathcal{W}^{-1} \mathcal{M}_\mathbb{C} \mathcal{W}$. Substituting this in \eqref{app1},
\begin{equation}
\left( \mathcal{W}^{-1} \mathcal{M}_\mathbb{C} \mathcal{W} \right)^{-1} = \mathcal{J}^{-1} \left( \mathcal{W}^{-1} \mathcal{M}_\mathbb{C} \mathcal{W} \right)^T \mathcal{J} \quad \Longrightarrow \quad \mathcal{M}_\mathbb{C}^{-1} = \mathcal{J}_\mathbb{C} \mathcal{M}_\mathbb{C}^T \mathcal{J}_\mathbb{C} \, ,
\end{equation}
where we have used $\mathcal{W}^{-1} = -\mathcal{W}^T$ and $\mathcal{J}^{-1} = -\mathcal{J}$. Thus,
\begin{equation}
\begin{pmatrix}
\Lambda & \Gamma \\
\Gamma^* & \Lambda^*
\end{pmatrix}^{-1}
=-
\begin{pmatrix}
1 & 0 \\
0 & -1
\end{pmatrix}
\begin{pmatrix}
\Lambda^T & \Gamma^\dagger \\
\Gamma^T & \Lambda^\dagger
\end{pmatrix}
\begin{pmatrix}
1 & 0 \\
0 & -1
\end{pmatrix}
=
\begin{pmatrix}
\Lambda^T & -\Gamma^\dagger \\
-\Gamma^T & \Lambda^\dagger
\end{pmatrix} \, .
\end{equation}

Now, writing $\mathcal{M}_\mathbb{C}^{-1}\mathcal{M}_\mathbb{C} = I$ explicitly, we arrive at several useful properties of $\Gamma$ and $\Lambda$, including $\Lambda^T \Lambda - \Gamma^\dagger \Gamma^* = I$. Applying this equality to an arbitrary vector $u \in \mathbb{C}^{2n}$,
\begin{equation}
\vert \Lambda u \vert^2 = \vert \Gamma u \vert^2 + |u|^2 \quad \Longrightarrow \quad \Arrowvert \Lambda \Arrowvert \geq 1 \, , 
\end{equation}
with $\Arrowvert \Lambda \Arrowvert = \sup_u | \Lambda u |/|u|$. It also follows is that $\Arrowvert \Lambda \Arrowvert \geq 1$. $\qquad \square$

Note that, by relying exclusively on the symplecticity of $\mathcal{M}$, this proof is valid regardless of whether $\mathcal{M}$ is a function of any parameter, including obviously time and phase-space points. For the original see \cite{FollandBook}.

\section{Compositions with the $\delta$-distribution}\label{App:B}

We limit ourselves to 1-dimensional spaces for brevity, but generalizations are trivial and can be found in \emph{e.g.}~\cite{SchwartzBook}. 

For $x \in \mathbb{R}$, Dirac's $\delta$ distribution is defined as the generalized functions fulfilling 
\begin{equation}
\int_\mathbb{R} dx \, \delta(x) \phi(x) = \phi(0) \, ,
\end{equation}
for all continuously differentiable test functions $\phi$. Its composition $\delta \circ f$ can be effortlessly worked out using a change of variables $f(x) \mapsto u$:
\begin{eqnarray}
	\int_\mathbb{R} dx \, (\delta \circ f)(x) \phi(x) &= \int_{f(\mathbb{R})} d[f^{-1}(u)] \, \delta(u) (\phi \circ f^{-1})(u) \\
	&= \int_{f(\mathbb{R})} du \, \delta(u) \left[ \frac{(\phi \circ f^{-1})(u)}{| (f' \circ f^{-1})(u)|} \right] \\
	&= \sum_{\ker(f)} \frac{\phi(x)}{|f'(x)|} \, , \label{B}
\end{eqnarray}
where the sum runs over the kernel of $f$, \emph{i.e.}~over all $x^{(i)}$ such that $f(x^{(i)}) = 0$. It is clear that this identification is only valid if $f'(x) \neq 0$. If we pick $\phi(x) = \sqrt{|f'(x)|}$ (which is also not continuously differentiable at the origin), we see that \eqref{B} becomes
\begin{align}
\int dx \, (\delta \circ f)(x) \sqrt{|f'(x)|} = \sum_{\ker(f)} \sqrt{\frac{1}{|f'(x)|}} \, .
\end{align}
Identifying $f \leftrightarrow x'(q,p;t) - q'$ and $x \leftrightarrow p$, where $x'(q,p;t)$ is a final position evolved by the hamiltonian flow and $p$ is the initial momentum, the condition above reads
\begin{align}
\int dp \, \delta \left[ x'(q,p;t) - q' \right] \left\vert \frac{\partial x'(q,p;t)}{\partial p} \right\vert^\frac{1}{2} = \sum_{p} \left\vert \frac{\partial x'(q,p;t)}{\partial p} \right\vert^{-\frac{1}{2}} \, ,
\end{align}
such that the left hand side is just an integral form of the root-search \cite{HellerBook}. If we integrate both sides with respect to $x'(q,p;t)$, the result is precisely Miller's trick \eqref{ivrq} in a 1-dimensional setting, with the filtering of final trajectories happening as a function of the initial momentum. We can run over any number of caustics in integration, since the area under the curve is asymptotically finite, as long as we do not end on them. If we do, it's just a matter of redefining the integration domain by associating the value $0$ to caustics, or deviating from the landed caustic by an infinitesimal value. As integration is blind to sets of zero measure, this either does not impact the result or sums an infinitesimal value to the final integral.

All the arguments above can be reformulated in the complex case, for which we then choose $f \leftrightarrow \gamma'(\zeta,\zeta^*;t) - \zeta'$ and $x \leftrightarrow \zeta$. The preservation of real orientation by complex determinants implies that the absolute values in the jacobians are not necessary, leading directly to \eqref{ivrz}. There is also no need to worry about caustics, since the complex pre-factor $\Lambda$ is non-singular (see \ref{App:A}).

\section{Implementing the root-search}\label{App:C}

We again restrict ourselves to the 1-dimensional case, as in Sec.~\ref{sec:kerr}. 

The vV-G propagator requires root-searching, \emph{i.e.}~in order to obtain the matrix element $\langle q' | \widehat{U}(t) | q \rangle$ we must find all initial momenta that link the classical trajectories $(q,p)$ at $t=0$ and $(q',p')$ at $t=t$. Evidently, the final momentum is not important in the process.

Now, regardless of whether or not the system at hand has an analytical solution, the process can be numerically implemented in the same fashion, enumerated below.

\begin{enumerate}
	\item Fix $q$, $q'$ and $t$;
	\item Sort an initial momentum $\widetilde{p}$ and use it to calculate a final position $Q'(q,\widetilde{p};t)$;
	\item Let $\epsilon>0$ such that $|Q'(q,\widetilde{p};t) - q'| < \epsilon$. The value attributed to $\epsilon$ is the threshold which a final position must overcome for its trajectory to be considered a solution to the root-search problem. Add $\widetilde{p}$ to the list of root momenta;
	\item The list of root momenta will consist of several momenta for any non-linear flow, and some of them will be very close to each other, since if $|Q'_1(q,\widetilde{p}_1;t) - Q'_2(q,\widetilde{p}_2;t)|<\epsilon$ and $\widetilde{p}_1$ is a root, then $\widetilde{p}_2$ will also be a root. This makes it necessary to define $\delta > 0$ such that root momenta that fulfill $|\widetilde{p}_2 - \widetilde{p}_1| < \delta$ must be considered to correspond to the same trajectory. 
\end{enumerate}

\begin{figure}
	\centering
	\subfigure[$\,$The geometry of root-searching]{
		\includegraphics[width=0.5\columnwidth, keepaspectratio]{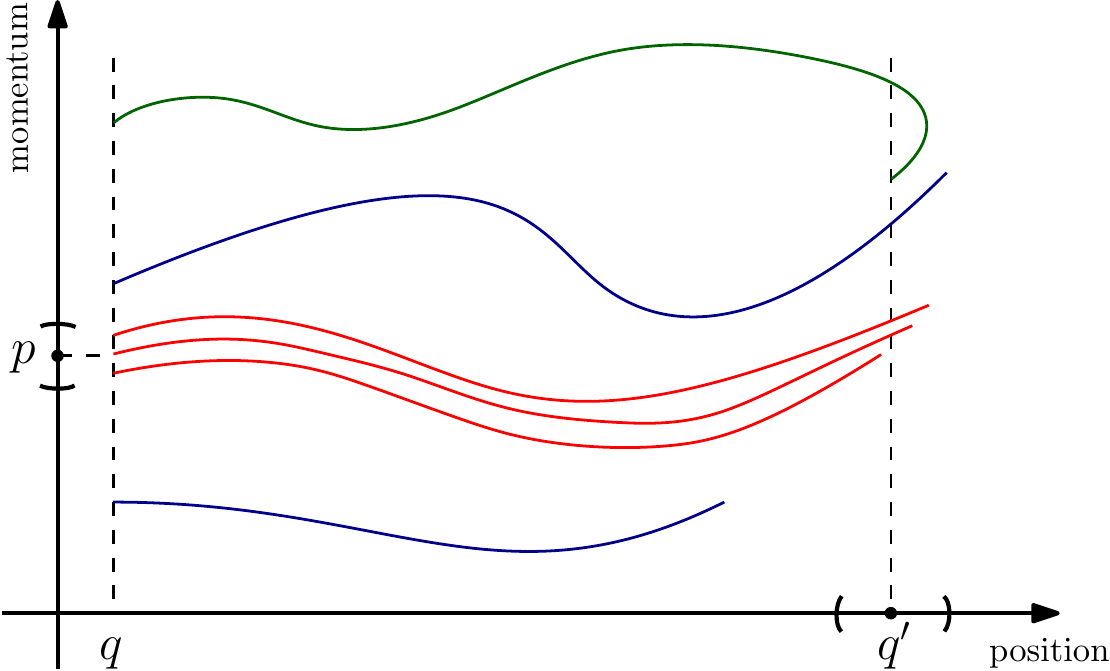}     
	}\hspace{1cm}
	\subfigure[$\,$Root-trajectories for Kerr]{
		\includegraphics[width=0.3\columnwidth, keepaspectratio]{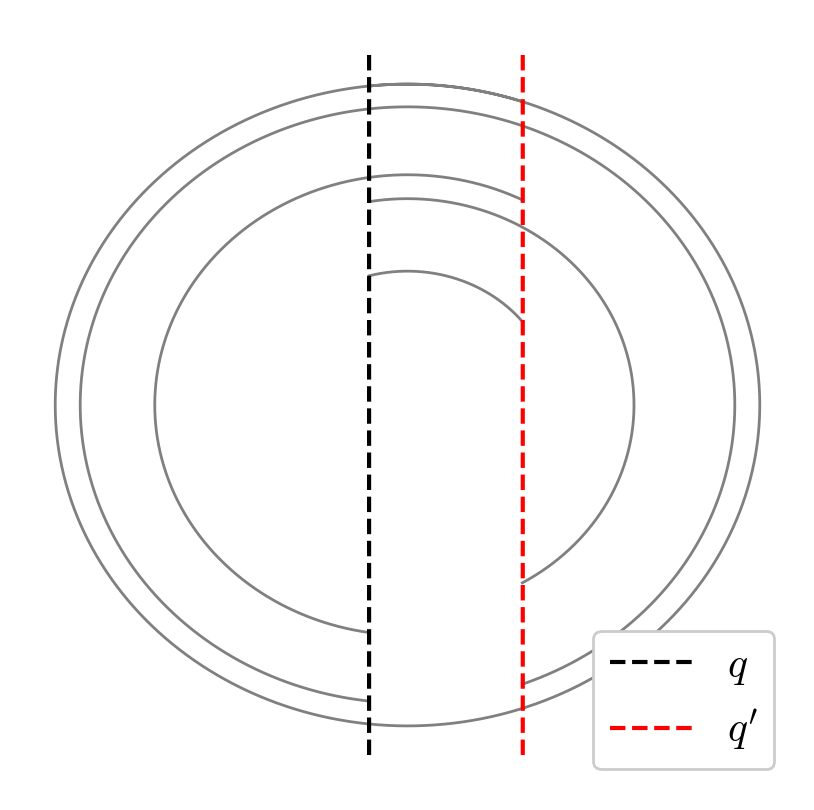}
	}
	\caption{(a) In this panel we can see that several root-trajectories, displayed in red, start at a fixed $q$ and end on a ball of radius $\epsilon$, centered at $q'$. These trajectories will all be selected by the root-search, but need to count as a single one. This is done \emph{via} a selection performed on the their initial momenta, which are all within a ball of radius $\delta$ centered at $p$. Since the lone green trajectory is outside this ball, its momentum counts as a new element to the root search and is not excluded, while the blue trajectories do not fulfill the root-search condition and are not selected. (b) Here we can see some true typical trajectories selected by the root-search. Notice the outermost one leaves $q$ and arrives at $q'$ after performing more than a complete period.}
	\label{fig:trajs}
\end{figure}

The procedure above requires sampling over grids of initial momenta to obtain each component of the vV-G propagator, but this process is actually faster than two dimensional integration for a single degree of freedom. The process, however, scales quite poorly with dimension, and then integral methods such as IVRs become more efficient. It must be kept in mind that the careless accounting for multiplicity in step 4.~is only possible because the trajectories in the Kerr system have monotonically increasing frequencies, such that true root-trajectories are never too close. Another important point if that setting $\epsilon$ smaller than the grid spacing we are using for $q$ and $q'$ has no impact on the results, since errors of this order are already present in the numerical implementation of the flow itself. The geometry of root-searching, together with its corresponding roots for the Kerr system, are displayed in Fig.~\ref{fig:trajs}.

\section*{References}
\bibliographystyle{unsrt}	 
\bibliography{HKvVG.bib}

\end{document}